\documentclass[pra,reprint,showpacs,showkeys,groupedaddress,nofootinbib,floatfix]{revtex4-1}
\usepackage{amsmath,amssymb,amsfonts}
\usepackage{graphicx}
\usepackage{txfonts}
\usepackage{epstopdf}
\usepackage[T1]{fontenc}

\usepackage[unicode]{hyperref}
\hypersetup{
    unicode=true,                                           
    a4paper=true,
    plainpages=false,
    pdffitwindow=true,                                      
    colorlinks=true,                                        
    linkcolor=blue,                                        
    citecolor=blue,                                         
    filecolor=blue,                                        
    urlcolor=blue                                          
}
\newcommand{\bra}[1]{\langle #1\vert}
\newcommand{\ket}[1]{\vert #1\rangle}
\newcommand{\bket}[2]{\langle #1\vert#2\rangle}
\newcommand{\ev}[1]{\langle #1 \rangle}

\newcommand{\pdsq}[1]{\frac{\partial^2}{\partial #1^2}}
\newcommand{\refeq}[1]{Eq.~(\ref{#1})}

\newcommand{\reffig}[1]{Fig.~\ref{#1}}

\begin{document}

\title{Collision dynamics and entanglement generation of two initially independent and indistinguishable boson pairs in one-dimensional harmonic confinement}
\author{David I. H. Holdaway}
\email{d.i.h.holdaway@dur.ac.uk}
\author{Christoph Weiss}
\author{Simon A. Gardiner}

  \affiliation{Joint Quantum Centre (JQC) Durham--Newcastle, Department of Physics, Durham University, Durham DH1 3LE, United Kingdom}

\date {\today}

 \begin{abstract}
We investigate finite number effects in collisions between two states of an initially well known number of identical bosons with contact interactions, oscillating in the presence of harmonic confinement in one dimension.  We investigate two $N/2$ (interacting) ground states, which are initially displaced from the trap center, and the effects of varying interaction strength.  The numerics focus on the simplest case of $N=4$.    
In the non-interacting case, such a system would display periodic oscillation with a half harmonic oscillator period (due to the left-right symmetry). With the addition contact interactions between the bosons, collisions generate entanglement between each of the states and distribute energy into other modes of the oscillator.  
We study the system numerically via an exact diagonalization of the Hamiltonian with a finite basis, investigating left/right number uncertainty as our primary measure of entanglement.  Additionally we study the time-evolution and equilibration of the single-body von Neumann entropy for 
both the attractive and repulsive cases.  We identify parameter regimes for which attractive interactions create qualitatively different behavior to repulsive interactions, due to the presence of bound states (quantum solitons) and explain the processes behind this.  
\end{abstract}


\maketitle

\section{Introduction}

Dilute gases of Alkali atoms have proved a powerful tool for the experimental investigation of quantum mechanical phenomena, from the level of single atom physics up to to mesoscopic levels via the creation of Bose-Einstein condensates (BECs)~\cite{AndersonEnsher1995,DavisMewes1995}.  Much of the interest stems from the ability to experimentally realize many theoretically interesting potentials, such as optical lattices~\cite{Bloch2005}, double well potentials~\cite{SchummEtAl05} and periodic kicking~\cite{RyuPhillips2006}, with the ability to control the effective dimensionality and interaction strength via Feshbach resonances.  Another interesting property is the ability to support both bright and dark solitons~\cite{MishmashCarr2009,SinhaJoachim2006}. 

Experimentally it is possibly to tune $s$-wave scattering lengths to both positive and negative values~\cite{BradleyHulet1997, KhaykovichSchreck2002,StreckerPartridge2002}.  However, beyond a certain critical number (which is dependent on the trapping configuration and scattering length), the negative scattering length (attractively interacting) systems are unstable to collapse~\cite{DoddBurnett1996,CornishThompson2006,BillamWrathmall2011b,ParkerCornish2007}.  If trapping potentials are present in two spatial dimensions, attractive condensates can exhibit self trapping, i.e., localization (at least in terms of pair correlation functions) in a direction free of external potentials. In quasi one-dimensional (1D) geometries, attractive BEC's form bright matter-wave solitons with particle-like dynamics for the center of mass~\cite{MartinAdams2008}.  Parameter regimes for which systems are quasi-1D have been investigated via variational techniques\cite{BillamWrathmall2011b}, along with effective potential approximations to deal with residual 3D effects~\cite{SalasnichParola2002}, leading to higher order effective non-linearities.  In addition to this, bright gap solitons~\cite{EiermannOberthaler2004} have been created from repulsive atoms in optical lattices by exploiting anomalous dispersion to give the atoms a negative effective mass.

Negative scattering lengths also give interesting possibilities in double-well and lattice physics. Repulsive interactions between atoms are known to give rise to the famous Mott insulator state~\cite{Spielman07}, with a near definite atom number per lattice site.  If one has a definite number of atoms per site, there is effectively a total uncertainty in relative phase between lattice sites and thus no phase coherence.  A measurement of relative phase should give totally random results and indeed this is what one finds when imaging the moment distribution of such a lattice, no distinguishable interference patterns.   
Attractive interactions could in theory be used to squeeze number statistics the opposite way, such that the ground state would tend to a superposition of a quantum soliton ($N$ atom bound state) delocalized over every lattice site.  When only two sites are present, such a state is referred to a NOON state~\cite{WildfeuerDowling2007}, 
which is useful for non shot noise limited interferometry~\cite{Lucke2011}. However, systems where the ground state is such a superposition are known to be extremely unstable to temperature, as phase differences between the two sites have almost no energy cost, thus typically replacing quantum uncertainty with statistical uncertainty.  It is therefore preferable to create such states dynamically, for example by splitting a moving quantum soliton~\cite{gertjerenken2012nonlocal,GertjerenkenWeiss2012scattering}.  

Any closed quantum system with no decoherence effects will be described by a wavefunction that will evolve deterministically. As such the wavefunction at any point in time $\vert \psi(t) \rangle$ maps back to a unique $\vert \psi(0) \rangle$. 
Recent experiments have shown great possibility to observe this deterministic behavior in systems with a small number of cold atoms~\cite{SerwaneJochim2011,ZurnJochim2012}, with dynamics that can be analytically calculated and with precise tuning available in the scattering length and confinement potentials.  Strongly correlated effects and quantum superpositions are generally much easier to achieve in few-body systems.  
Despite this one can still envisage collective properties (such as expectation values of operators) of a time-dependent finite system tending to constant values when averaged over reasonable timescales, or relaxation of local operators, as shown in~\cite{CramerOsborne2008}.  Non-integrable systems, upon coupling to another larger system, usually tend to an equilibrium configuration at long times, independent of the initial state of either system (except for the total energy); however recent theoretical observations have thrown doubt on this~\cite{GogolinEisert2011}.  Additionally, when two coupled systems contain a similar number of elements the situation is less clear still. Our system is non-integrable and contains two initially independent subsystems of the same size; hence, we are interested to what extent equilibration occurs or where it is resisted.   
Quantum systems, for example atoms populating sites in an optical lattice~\cite{GreinerBloch2009} are known to show partial revivals of the initial state in time, but are generally observed to show weaker revivals as time progresses in an apparent damping; we are interested in whether certain measures, specifically the number to the left and right of are trap and the single body von Neumann entropy, tend to constant values when averaged over sufficient timescales.  

This paper is organized as follows:
Section~\ref{sec:sys} introduces the one dimensional Hamiltonian and the unit rescaling to harmonic oscillator lengths, constant throughout the paper.  Next, the initial condition is introduced, with specific cases of interest mentioned. 
Section~\ref{sec:entangle} discusses observables and measures of entanglement that we will use to investigate the system, included the variation in the number to either side of the trap center and the single-body von Neumann entropy.  
Section~\ref{sec:int_sys} begins an analytic investigation of the system, focusing on the mechanisms by which interactions modify the dynamics of each displaced state and generate entanglement.  
Section~\ref{sec:experimental} discusses a possible experimental realization of the system, using ultracold atoms in an optical lattice, with parameters discussed for Cesium.    
Section~\ref{sec:numeric_meth} contains a brief description of the numerical method, based on exact diagonalization. 
Section~\ref{sec:results} presents numerically obtained results for the evolution of our observables and entanglement measures in the system. 
Section~\ref{sec:conc} summarizes and concludes. 

\section{System \label{sec:sys}}
\subsection{Hamiltonian and unit rescaling}
We consider an effective one-dimensional (1D) system [taken to be reduced from a three-dimensional (3D) configuration where the radial degrees of freedom are strongly confined by a harmonic trapping potential] of structureless bosons subject to attractive or repulsive contact interactions $V(|x_1-x_2|) = g_{\textrm{1D}} \delta(x_1-x_2)$, i.e., a Lieb--Liniger--(McGuire) gas~\cite{HoldawayWeiss2012,McGuire1964,LiebLiniger1963}, with the addition of an axial harmonic confining potential. In second-quantized form, this can be described by the following Hamiltonian:\begin{multline}
 \hat{H} =  \int  \mathop{dx}  \hat{\Psi}^{\dagger}(x)\left(-\frac{\hbar^2}{2M}\pdsq{x}+\frac{M\omega_{x}^2 x^2}{2}
 \right)\hat{\Psi}(x) 
\\ 
+ \frac{g_{\textrm{1D}}}{2}  \int  dx \hat{\Psi}^{\dagger}(x)\hat{\Psi}^{\dagger}(x)\hat{\Psi}(x)\hat{\Psi}(x)\; ,
\label{Eq:SecondQuantizedH}
\end{multline}
where $M$ is the mass and $\omega_x$ the axial (angular) trapping frequency; assuming a radial trapping frequency of $\omega_{\rm r}$, the coupling parameter $g_{\textrm{1D}} = 2\hbar\omega_{\rm r}a_{s} $,  with $a_{s}$ the (3D) $s$-wave scattering length~\cite{PitaevskiiStringariBook2003,Olshanii1998}. A satisfactory condition for this Hamiltonian to be valid is $N \vert a_{s} \vert \ll \sqrt{\hbar/M\omega_{\rm r}}$ and $k_B T \ll \hbar \omega_r$, however it is likely to be still be valid for $k_B T \sim \hbar \omega_r$, i.e., as long as thermal excitations are unlikely to significantly populate radial modes. 

We use harmonic oscillator units (codified as $\hbar = \omega_{x} = M =1$), meaning that length is in units of $\sqrt{\hbar/M\omega_{x}}$, time in units of $1/\omega_{x}$, and energy in units of $\hbar \omega_{x}$; a harmonic oscillator period is then $2 \pi$.  The Hamiltonian rescales to
\begin{equation}
\hat{H} =  \int \mathop{dx} \hat{\Psi}^{\dagger}(x)\left[-\frac{1}{2}
\pdsq{x}+\frac{x^2}{2}
+ \frac{g}{2}  
\hat{\Psi}^{\dagger}(x)\hat{\Psi}(x)\right]\hat{\Psi}(x) \; ,
\label{Eq:ScaledSecondQuantizedH}
\end{equation}
where $g=g_{1{\rm D}} \sqrt{M/\hbar^3 \omega_x}$ is the new dimensionless coupling parameter.\footnote{This relates to the parameter $\gamma$ of \cite{HoldawayWeiss2012} through $\gamma = [g(N-1)]^{-2}$.}  In first quantization we can express this same Hamiltonian (for $N$ particles) as
\begin{equation}
H(\vec{x}) = \sum_{k=1}^N \left( -\frac{1}{2} \pdsq{x_{k}} + \frac{x_{k}^2 }{2} \right)
 + g \sum_{k=2}^{N} \sum_{j =1}^{k-1} \delta(x_{k}-x_{j}) \; ,
\label{eq:first_quant_ham}
\end{equation}
where $x_{k}$ are the coordinates of the individual particles (generally considered to be ultracold atoms), and $\vec{x}$ is a shorthand for the set of all $N$ coordinates $\{x_{1},x_{2},\ldots,x_{N}\}$.  As the external potential is harmonic,  $H(\vec{x})$ can be partitioned into two mutually commuting components~\cite{HoldawayWeiss2012}, one describing the center of mass (giving rise to the Kohn mode~\cite{BonitzBalzer2007}), and the other describing the remaining degrees of freedom.  This separation can be exploited computationally, as the center-of-mass dynamics are those of a simple harmonic oscillator and can therefore be described exactly, reducing the effective dimensionality of the computational problem to $N-1$.

\subsection{Initial condition}
\subsubsection{General $N$-body case}
We consider a highly non-mean-field-like initial condition, taking two $N/2$-atom ground states (for a given $g$), equally and oppositely displaced from the trap center by a distance $x_0$, and symmetrizing. The initial ($t=0$) wavefunction is then
\begin{equation}
\begin{split}
 \psi(\vec{x},0) = &\frac{B}{\sqrt{N!}} \sum_{\{\cal P \}} f^{(N/2)}(x_1-x_0,..,x_{N/2}-x_0) \\
 &\times f^{(N/2)}(x_{N/2+1}+x_0,...,x_N+x_0) \;,
\end{split}
\label{eq:initcon2}
\end{equation}
where $f^{(N/2)}(x_1,\ldots,x_{N/2})$ is the ground state for $N/2$ atoms (generally numerically determined) in the harmonic trap, $\{{\cal P }\}$ is the set of all permutations of $\vec{x}$, and $B$ is a normalizing factor.  Such an initial condition may be motivated by the idea of making two separate BECs and allowing them to collide within a harmonic trapping potential, or from rapidly modifying a Mott insulator state in an optical lattice (as we will discuss in Section \ref{sec:experimental}).  If the left and right components are well separated, i.e., the width of the atomic density distribution corresponding to $f^{(N/2)}$ is significantly less than $x_{0}$, then there is a well defined number of $N/2$ atoms either side of the trap, and left- and right-atoms are distinct by virtue of their position.  Furthermore, as the center-of-mass dynamics are decoupled~\cite{HoldawayWeiss2012} and straightforward to determine, the dynamics experienced by an initial condition such as $\psi$ can be readily extended to incorporate any initial condition for the center of mass, e.g., in particular, an overall oscillation about the trap center~\cite{MartinAdams2008}.

Conveniently, $\psi(\vec{x},t)$ is in the ground state of the center-of-mass component of $H(\vec{x})$.  To show this, we first define (unnormalized) Jacobi coordinates, consisting of the center-of-mass coordinate
\begin{equation}
x_{\textrm{C}(N)} = \frac{1}{N}\sum_{k=1}^N x_k,
\label{eq:xcdef}
\end{equation}
and $N-1$ further independent coordinates 
\begin{equation}
\xi_{k} =  x_{k} - \frac{1}{k-1}\sum_{j=1}^{k-1} x_{j} \;,
\label{Eq:JacobiCoordinates}
\end{equation}
where $k\in\{2,3,\ldots,N\}$.  Using the Jacobi coordinates for $N/2$ particles, we can partition the $N/2$-particle ground state into center-of-mass-dependent and independent components: $f^{(N/2)}(x_{1}, \ldots, x_{N/2}) = \varphi(\xi_{2},\ldots,\xi_{N/2})\mathop{\mathrm{e}^{-N x_{\mathrm{C}(N/2)}^{2}/4}}$. 
Substituting in \refeq{eq:sumxsqid2}, we can then define $\tilde{f}^{(N/2)}$ through
\begin{align}
f^{(N/2)}(x_{1}, \ldots , x_{N/2}) = &\varphi(\xi_{2},\ldots,\xi_{N/2})
\mathop{\mathrm{e}^{\sum_{k=2}^{N/2}[(k-1)/2k]\xi_{k}^{2}-\sum_{k=1}^{N/2}x_{k}^{2}/2}}
\nonumber \\
= & 
\tilde{f}^{(N/2)}(x_{1}, \ldots, x_{N/2})
\mathop{\mathrm{e}^{-\sum_{k=1}^{N/2}x_{k}^{2}/2}},
\end{align}
where $\tilde{f}^{(N/2)}$ (as it can also be written as a function of $\{\xi_{2},\xi_{3},\ldots,\xi_{N/2}\}$ only) is clearly independent of $x_{\mathrm{C}(N/2)}$.  Within an expanded set of $N$ coordinates, $\tilde{f}^{(N/2)}(x_{1}, \ldots, x_{N/2})$ is also clearly independent of $x_{\mathrm{C}(N)}$, as is (by symmetry) $\tilde{f}^{(N/2)}(x_{N/2+1}, \ldots, x_{N})$.  Noting further that displacement by $x_{0}$ will not affect that part of $f^{(N/2)}$ independent of the center-of-mass coordinate, then for the identity permutation of $\psi$
\begin{multline}
  f^{(N/2)}(x_1-x_0,\ldots,x_{N/2}-x_0) f^{(N/2)}(x_{N/2+1}+x_0,\ldots,x_N+x_0) 
\\
=\tilde{f}^{(N/2)}(x_1,\dots,x_{N/2})\tilde{f}^{(N/2)}(x_{N/2+1},\ldots,x_{N}) 
\\
\times 
\mathrm{e}^{-\sum_{k=1}^N x_k^2/2 - x_0\left[
\sum_{k=N/2 +1}^{N}  x_k
-\sum_{k=1}^{N/2} x_k 
\right] - N x_0^2/2}\;.
\end{multline}
By the identities \refeq{eq:sumxsqid2} and \refeq{eq:secondJidentity}, the exponential reduces to 
$\mathrm{e}^{-Nx_{\textrm{C}(N)}^2/2}
\mathrm{e}^{-N\left(\sum_{k=N/2+1}^{N} \xi_{k}/k-x_0^2/2\right)}
$, i.e., a term proportional to the center of mass ground state multiplied by a function of independent Jacobi coordinates.  The identity permutation of $\psi$ can thus be written as a product of the center of mass ground state and a function of the other independent coordinates. This separation off of the center of mass ground state occurs for every permutation of the coordinates $x_{k}$, and so we conclude that the center-of-mass component of $\psi$ is indeed in the ground state.

Taking a slightly different initial condition, when one combines ground states from two trapping potentials which are not equal to the final potential (with, e.g., tighter harmonic trapping), will introduce a breathing motion, which can still be considered separately from the remaining dynamics.  It is also significant to note that the kind of initial condition we consider does not have a well defined relative phase between the left and right components
~\cite{Peggbarnett1997}.  If a relative number uncertainty between left and right were to develop then this would no longer be the case, and a meaningful relative phase could in principle be extracted.

\subsubsection{Time evolution for the non-interacting case \label{sec:NItimeev}}
If we take the case where $g=0$,  we can express the full time dependent wavefunction [which we label $\psi_{0}(\vec{x},t)$] analytically, as a symmetrizing product of two $N/2$-atom product states 
\begin{equation}
\psi_{0}(\vec{x},t) = \frac{B_{0}}{\sqrt{N!}} \sum_{\{\cal P \}}  \prod_{k=1}^{N/2} \phi(x_k,-x_0,t)
 \prod_{j=N/2+1}^{N} \phi(x_j,x_0,t) \;.
 \label{eq:initconNI}
\end{equation}
Here $\phi(x,\pm x_0,0)$ is a Gaussian displaced by $\pm x_0$ from the trap center, and~\cite{GerryKnight2004,MartinAdams2008} 
\begin{equation}
\begin{split}
\phi(x,x_0,t) = &\left(\frac{1}{\pi}\right)^{1/4} \exp\left(-\frac{[x-x_0 \cos(t)]^2}{2}\right)  \\
&\times \exp(i[t/2 - x_0 \cos(t) x + x_0 \sin(2t)/4])  \;,
\end{split}
\label{eq:osccoh}
\end{equation}
corresponding to an energy-per-particle of $E = (x_0^2+1)/2$, and where the normalization constant $B_{0} =1 + {\cal O}(\mathrm{e}^{-2x_0^2} )$.

\subsubsection{$N=4$ special case}
  If $N=4$, the $f^{(2)}$ appearing in \refeq{eq:initcon2} are known analytically~\cite{BuschEnglert1998,SowinskiBrewczyk2010}, and if $g < 0$ may, for  sufficiently large $g$ and $x_{0}$, be considered to be bound-state  dimers, held within an overall harmonic trapping potential.  The general form is given by
\begin{equation}
\begin{split}
f^{(2)}(x_{1} , x_{2}) &= 
\mathcal{N}
U\left(-\nu,1/2,\frac{[x_{1}-x_{2}]^2}{2}\right) 
\mathop{\mathrm{e}^{-x_{1}^{2}/2}}
\mathop{\mathrm{e}^{-x_{2}^{2}/2}}
\\
&= 
\mathcal{N}
U (-\nu,1/2,\xi_2^2/2 ) 
\mathop{\mathrm{e}^{-\xi_{2}^{2}/4}}
\mathop{\mathrm{e}^{-x_{\mathrm{C}(2)}^{2}}}
\;,
\end{split}
\label{eq:twobodygs}   
\end{equation}
with $U$ Tricomi's confluent hypergeometric function, $\mathcal{N}$ a normalization constant, and $\nu$ the effective quantum number (equal to zero for $g=0$), as determined by the transcendental equation $\Gamma(1/2-\nu)/\Gamma(-\nu) = -g/2^{3/2}$. This state has an energy of $2 \nu +1$, where there is a contribution of $1/2$ due to the center of mass.  
Equation (\ref{eq:twobodygs}) can then be inserted into the initial condition
\begin{equation}
\begin{split}
\psi(x_1,x_2,x_3,x_4,0) 
= & \frac{B}{\sqrt{4!}} \sum_{\{\cal P \}} 
f^{(2)}(x_1-x_0,x_2-x_0) 
\\ &\times 
f^{(2)}(x_3+x_0,x_4+x_0) \;,
\end{split}
\label{eq:initcon}
\end{equation}
where $\{\cal P \}$ is the set of all $4!$ permutations of $\{x_1,x_2,x_3,x_4\}$. Note that, as $f^{(2)}(x_{1},x_{2}) = f^{(2)}(x_{2},x_{1})$, the number of distinct permutations actually reduces to $4!/2!2! = 6$.

\section{Observables and measures of entanglement\label{sec:entangle}}

\subsection{Left/right number \label{sec:numlr}}

For our system, one useful measure to track the generation of entanglement is the variance in particle number to the left and right of the system's center of mass (which we will generally consider to be fixed at the origin). The initial condition we consider has $N/2$ atoms to either side with essentially no possibility of, say, $N/2+1$ to the right and $N/2-1$ to the left (probabilities for measuring such unequal partitionings decrease Gaussianly with the initial separation).  Hence, the left- and right-particle-number-variance will initially be zero.  As the left- and right-particles approach and collide, all number partitionings become possible, and so this measure is only informative when the particle density at the location of the center of mass is small.

We define a ``number-to-the-right'' operator
\begin{equation}
 \hat{N}_R = \int_{0}^{\infty} dx \hat{\Psi}^{\dagger} (x) \hat{\Psi}(x) \; ,
 \label{eq:numlr}
\end{equation}
[or in first quantization $\sum_{k=1}^N \Theta(x_k)$, where $\Theta$ is the Heaviside step function]; imaging one side of the trap would correspond to a projective measurement into the eigenstates of this operator, as is discussed in section \ref{sec:experimental}. The expectation value of $\hat{N}_R$ is the mean number of particles on the right-hand-side --- as the system is parity preserving, $\ev{\hat{N}_R} = N/2$ for all time for the initial conditions we consider.  

The more informative number-to-the-right variance is
\begin{equation}
\Delta N_R = \ev{\hat{N}_R^2} - \ev{\hat{N}_R}^2 \;,
\end{equation}
which, for our initial condition of two well separated left-and-right components of definite number, should be $\approx 0$.  From \refeq{eq:DeltaNR}, the variance for a product state $\psi(\vec{x}) = \prod_{k=1}^{N} \phi(x_k)$ [symmetric about the trap center so that $\ev{\hat{N}_R}=N/2$] is 
\begin{equation}
   \Delta_{P}N_{R}
   = \ev{\hat{N}_R} (1-\ev{\hat{N}_R}/N ) = N/4\; ,
\label{eq:DeltaNRmain}
\end{equation}
which evaluates to unity if $N=4$ (this is however the same as a symmetric superposition of one and three atoms to the right /left).  It can also be shown (Appendix \ref{app:AnalyticLRvar}) that for the case of $N=4$ and no interactions ($g=0$) [given by \refeq{eq:initconNI}], this variance evolves as
\begin{align}
   \Delta_{P}N_{R}
= 1-\mbox{erf}^2(x_0 \cos(t)) + {\cal O}(\mathrm{e}^{-2 x_0^2}) \; , 
\label{eq:DeltaNRnistate}
\end{align}
with erf the error function.\footnote{Satisfying $\mbox{erf}(0) = 0$ and $\mbox{erf}(\pm x) \to \pm [1 - \exp(-x^2)/(\sqrt{\pi}x) ]$ as $x\to \infty$.}  Hence, we have a function with period $T=\pi$, which is equal to unity when $t = (n+1/2)\pi$ and vanishingly small in $x_0$ when $t = n\pi$. 

In general our wavefunction is not an eigenstate of $\hat{N}_{R}$, and contains components of different  $\hat{N}_{R}$ eigenstates (for some given overall $N$, meaning that an additional specification of number-to-the-left operator eigenstates is not necessary).

One can however calculate expectation values of operators defined over restricted regions of  state space, specific to having exactly $n$ (of $N$) atoms to the right of the trap center. An expectation value for an operator $\hat{O}$ defined in this region is then
\begin{equation}
 \ev{\hat{O}}_{n,N-n}
 = \frac{\int_0^{\infty} \mathop{dx_1\ldots dx_n} 
 \int_{-\infty}^0  \mathop{dx_{n+1}\dots dx_{N}} \psi^*(\vec{x}) \mathop{O(\vec{x})} \psi(\vec{x})}
 {\int_0^{\infty} \mathop{dx_1\ldots dx_n} \int_{-\infty}^0  \mathop{dx_{n+1}\ldots dx_{N}} \vert \psi(\vec{x})\vert^2}  \; . 
\label{eq:restricted_ev}
\end{equation}
This is equivalent to taking the usual expectation value over a new wavefunction $\psi_{n,N-n}(\vec{x})$ defined by
\begin{equation}
\psi_{n,N-n}(\vec{x}) = \frac{1}
{\sqrt{P_{n,N-n}}}
\psi(\vec{x})\sum_{\cal P} 
\prod_{k=1}^n \Theta(x_k) 
\prod_{j=n+1}^N \Theta(-x_j) \; , 
\label{eq:restricted_wf}
\end{equation}
where ${\cal P}$ is the set of all unique permutations, of which there are $N!/n!(N-n)!$, and $P_{n,N-n}$ is a normalizing factor, giving the probability of finding $n$ of $N$ atoms to the right (or equivalently $N-n$ to the left) of the trap center.  Each such wavefunction is an eigenstate of $\hat{N}_{R}$, 
with eigenvalue $n$.  In principle one can partition the Hilbert space in such a way that it is the tensor product of a subspace describing only how many particles are to the left/right of the trap center, and a subspace describing all other relevant properties of the system state.  We may denote the set of eigenstates of $\hat{N}_{R}$ spanning this ``number subspace'' by  $\{ \ket{N-n,n} \}$, such that
\begin{equation}
 \hat{N}_R \ket{N-n,n} = n \ket{N-n,n} \; .
\label{eq:resregdef}
\end{equation}
We can also study expectation values of a distance-to-the-right operator $\int_0^{\infty} \mathop{dx} x \hat{\Psi}^{\dagger}(x) \hat{\Psi}(x)$ [$\sum_{k=1}^{N} \Theta(x_k) x_k$ in first quantization] and associated higher-order moments, which will trace particle-like tracks (and widths around them) for state components of different right-hand number $n$.

\subsection{von Neumann entropy and relaxation}

Averaging over all individual particles results in the single-body density matrix
\begin{equation}
\rho(x,x',t) = \ev{\hat{\Psi}^{\dag}(x') \hat{\Psi}(x)} \;,
\label{eq:sbdm}
\end{equation}
which is normalized to the total particle number $N$.  From this, single-body properties of the many-body system may be determined, specifically the von Neumann entropy\footnote{This is sometimes referred to as the Invariant Correlation Entropy (ICE)~\cite{Sokolov1998}.}
\begin{equation}
 S_{\mathrm{VN}}(t) = -\mathop{\mathrm{Tr}}\left\{(\rho/N) \ln (\rho/N)\right\} \;.
\label{eq:VNE}
\end{equation}

Relaxation, in the sense of tending 
to states of higher entropy, is not present if the system is fully integrable, i.e., when $g=0$, or if the trapping is removed and the eigenstates are given by the Bethe ansatz~\cite{LiebLiniger1963}.  However, as the integrability is broken by the trapping, we expect some degree of thermalization due to (previously forbidden) mixing between states.  It is of interest to determine how such thermalization timescales vary with the interaction strength and initial separations.  

For a product state, $\rho$ has a single non-zero eigenvalue of value $N$, meaning $S_{\mathrm{VN}} \to 0$ (this is equivalent to a Bose--Einstein condensate being exactly described by a Gross--Pitaevskii wavefunction).  A larger value of $S_{\mathrm{VN}}$ indicates occupancy of multiple eigenstates of $\rho$, equivalent to population of non-condensate modes due to to thermal excitations, or to quantum or dynamical depletion~\cite{BillamGardiner2012,BillamMason2012}.

If the system equilibrates, $S_{\mathrm{VN}}$ will tend to a constant value.  As our initial conditions result in repeated collisions at the trap center,  the value of $S_{\mathrm{VN}}$ shows distinct oscillations that decay only slowly. We therefore also consider a time average over an oscillator period
\begin{equation}
 \bar{S}_{\mathrm{VN}}
 (t) = \frac{1}{2 \pi} \int^{t + 2\pi}_{t} \mathop{dt'} S_{\mathrm{VN}}(t') \; ,
\label{eq:VNEav}
\end{equation}
along with its variance
\begin{equation}
 \Delta \bar{S}_{\mathrm{VN}}
 (t) = \int^{t + 2\pi}_{t} \mathop{dt'} \left[\frac{S_{\mathrm{VN}}(t')}{2 \pi} 
 - \bar{S}_{\mathrm{VN}}
 (t)\right]^2 \; .
\label{eq:VNEvar}
\end{equation}
If $S_{\mathrm{VN}} (t)$ tends to a constant value, this will be shown by a relaxation of $\bar{S}_{\mathrm{VN}} (t)$ to a constant value, and  a relaxation of $\Delta \bar{S}_{\mathrm{VN}}(t)$ to $0$, with the relaxation of  $\bar{S}_{\mathrm{VN}} (t)$ tending to occur on a significantly faster time scale than that of $\Delta \bar{S}_{\mathrm{VN}}(t)$.

\section{Analysis of the interacting system \label{sec:int_sys}}
\subsection{Left--right separation of the Hamiltonian}

As our initial condition consists of left and right components which are well separated and therefore distinguishable, we can initially treat the left and right components separately. As these left and right clusters only interact for a short-time during collisions in the center (so long as they stay as distinct clusters), it makes sense to treat interactions between these clusters perturbatively at early times.  We therefore split the Hamiltonian into three, restricting the coordinates to the region $x_{1} \le x_{2} \le x_{3} \le x_{4}$, which is sufficient due to Bose symmetry. The three components are
\begin{equation}
\begin{split}
 H_{L}(x_{1},x_{2}) &= \sum_{k=1}^2 \left( -\frac{1}{2} \pdsq{x_{k}} + \frac{x_{k}^2 }{2} \right) + g  \delta(x_{2}-x_{1})  \; , \\
 H_{R}(x_{3},x_{4}) &= \sum_{k=3}^4 \left( -\frac{1}{2} \pdsq{x_{k}} + \frac{x_{k}^2 }{2} \right) + g \delta(x_{4}-x_{3})  \; , \\
 H_I(x_{2},x_{3}) &= g \delta(x_{3}-x_{2})  \; .
\end{split}
 \label{eq:split_ham}
\end{equation}
The reason only adjacent interaction terms [$\delta(x_k-x_j)$ with $k-j = 1$] remain is that the other terms constitute a set of zero measure in the region we are considering, i.e., $x_1 = x_2$ occurs infinitely more often than $x_1 = x_3$, which necessarily implies $x_2 = x_3$ and so is a set of lower dimensionality.  
As $[ \hat{H}_L,\hat{H}_R ] = 0$, if we neglect $\hat{H}_I$ our system can be described by a tensor product of the left and right components.\footnote{Commuting Hamiltonians imply 
$\exp(-i [\hat{H}_L+\hat{H}_R]t) \ket{\psi} = \exp(-i \hat{H}_L t)  \ket{\psi_L}  \exp(-i \hat{H}_R t)  \ket{\psi_R}$, i.e., the time evolution operator can be separated.}  Each Hamiltonian $\hat{H}_{L/R}$ can further be split into center-of-mass $\hat{H}_{L/R}^{(\rm C)}$ and relative parts $\hat{H}_{L/R}^{(\rm R)}$, generating the dynamics of the left and right center-of-mass and relative coordinates
[$x_{\mathrm{C}(L)} = (x_1 +x_2)/2$, $x_{\mathrm{C}(R)} = (x_3 +x_4)/2$, $x_{\mathrm{R}(L)} = x_2 -x_1$, and $x_{\mathrm{R}(R)} = x_4 -x_3$, respectively],
 which again mutually commute.

We consider the center-of-mass wavefunction of an $n$ atom cluster, which is a Gaussian displaced from the trap center by some value $X_n$.  Without the influence of $\hat{H}_I$ our system consists of two indistinguishable clusters (with internal degrees of freedom considered to be in the ground state) undergoing simple harmonic motion.  The primary reason for separating the Hamiltonian in this way is that our initial condition is in the ground state of $\hat{H}_{L/R}^{(\rm R)}$ and is a displaced ground state of $\hat{H}_{L/R}^{(\rm C)}$, hence any change to these wavefunctions is an excitation of the system.

\subsection{Perturbative introduction of $H_I$}
\subsubsection{Overview}
We consider the effect of introducing the Hamiltonian $H_I$, from \refeq{eq:split_ham}, to the system.  
We look at three notable effects:  
changes to the wavefunction describing the left/right separation of the clusters;  
changes to the internal degrees of freedom within the clusters to the left and right; 
and interactions transferring atoms from one side to the other, creating a symmetric superposition. 

\subsubsection{Inter-cluster wavefunction changes and pseudo-periodicity \label{subsubsec:inter-cluster}}
The center-of-mass wavefunctions of each side, described by $\hat{H}_{L}^{(\rm C)} + \hat{H}_{R}^{(\rm C)}$, can change, so long as the \textit{global} center-of-mass wavefunction remains constant.  Such changes lead to entanglement between the left and right clusters, to see this we note initially the two cluster wavefunction could be written as a product of left and right sides
\begin{align}
\psi_0(x_{\mathrm{C}(L)},x_{\mathrm{C}(R)}) \propto \mathrm{e}^{-[x_{\mathrm{C}(L)} - x_0]^2}\mathrm{e}^{-[x_{\mathrm{C}(R)} + x_0]^2} + \mbox{perm}  \;,
\end{align}
with ``perm'' denoting the permutation of $R$ and $L$.  This can be written in such a way as to explicitly separate the global center of mass:
\begin{align}
\label{eq:globalcomsep}
\psi_0(x_{\mathrm{C}(L)},x_{\mathrm{C}(R)})=\;  &\mathrm{e}^{-[x_{\mathrm{C}(L)}+x_{\mathrm{C}(R)}]^2/2}  \\
&\times \left\{\mathrm{e}^{-[x_{\mathrm{C}(L)}-x_{\mathrm{C}(R)} - 2x_0]^2/2}  + \mbox{perm} \right\} \nonumber \; .
\end{align}
The first term describes the global center-of-mass and is therefore fixed, the latter term, however,  will be modified by interactions.  Any such change (other than modifying $x_0$ or multiplying by $\exp(i p[x_{\mathrm{C}(L)}-x_{\mathrm{C}(R)}])$, which are simply rescalings of the initial position and kinetic energy, respectively) means there will be terms involving products of the form $x_{\mathrm{C}(L)} x_{\mathrm{C}(R)}$, 
such that the wavefunction cannot be separated, indicating entanglement between the left and right sides.  Such entanglement is notable in the context of solitons in free space, as integrability means collisions cannot create entanglement once the states are asymptotically separated, although higher order non-linearities can also lead to entanglement~\cite{LewensteinMalomed2009}.  
Additionally, during collisions with attractive (repulsive) interactions, each cluster will accelerate (decelerate), subsequently returning to near its initial velocity, leading to a pseudo-periodicity. 

\subsubsection{Intra-cluster wavefunction changes \label{subsubsec:intra-cluster}}
The internal degrees of freedom described by $\hat{H}_{L/R}^{(\mathrm{R})}$ are initially in the ground state. Interactions during collisions will introduce excitations, 
with the energy transferring from the center-of-mass energy of each cluster.  By conservation of energy this must reduce the amplitude of the oscillation.  
Attractive interactions will suppress such excitations, as the energy separation between ground and first (even parity) exited state is greater than the harmonic oscillator level spacing, whereas for repulsive interactions this gap will be smaller.  Note that when highly excited modes of the relative degrees of freedom  $x_{\mathrm{R}(L)}, x_{\mathrm{R}(R)}$ are populated, these will always have a significant occupation for both $L$ and $R$.  One expects a qualitative difference in behavior between the attractive and repulsive cases to occur when the change between the first and second relative excited states differs by an amount of order unity in harmonic oscillator units $[\hbar\omega_{x}]$.  We note that for strongly attractive interactions,\footnote{In this regime energies scale as $-g^2 n(n^2-1)/2$ for an $n$ atom ground state~\cite{McGuire1964}.}, when $x_0 < -g/4$ 
there is not enough energy to break the bound-state clusters, making the relative degrees of freedom effectively inaccessible, but this is beyond the scope of the present paper. 

\subsubsection{Left/right atom transfer \label{subsubsec:transfer}}
Finally, the interactions can transfer an atom from one side to the other, mixing to a set of states with a symmetric superposition of three and one atoms at either side of the trap (and, ultimately, back from this to the original state).  There cannot be significant transfer to a state where there is a cluster of four atoms in the ground state (apart from the center-of-mass degree of freedom) on one side and zero on the side, due to the invariance of the center-of-mass wavefunction, unless the state has all four atoms directly at the trap center.  The state satisfying this condition is the ground state of the system, and so the only possible population is that there at $t=0$. Note that excited states of this four atom cluster do make up parts of the oscillating cluster states, it is simply a different basis to consider the system in terms of. 

\begin{figure}
\begin{center}
\includegraphics[width=\linewidth]{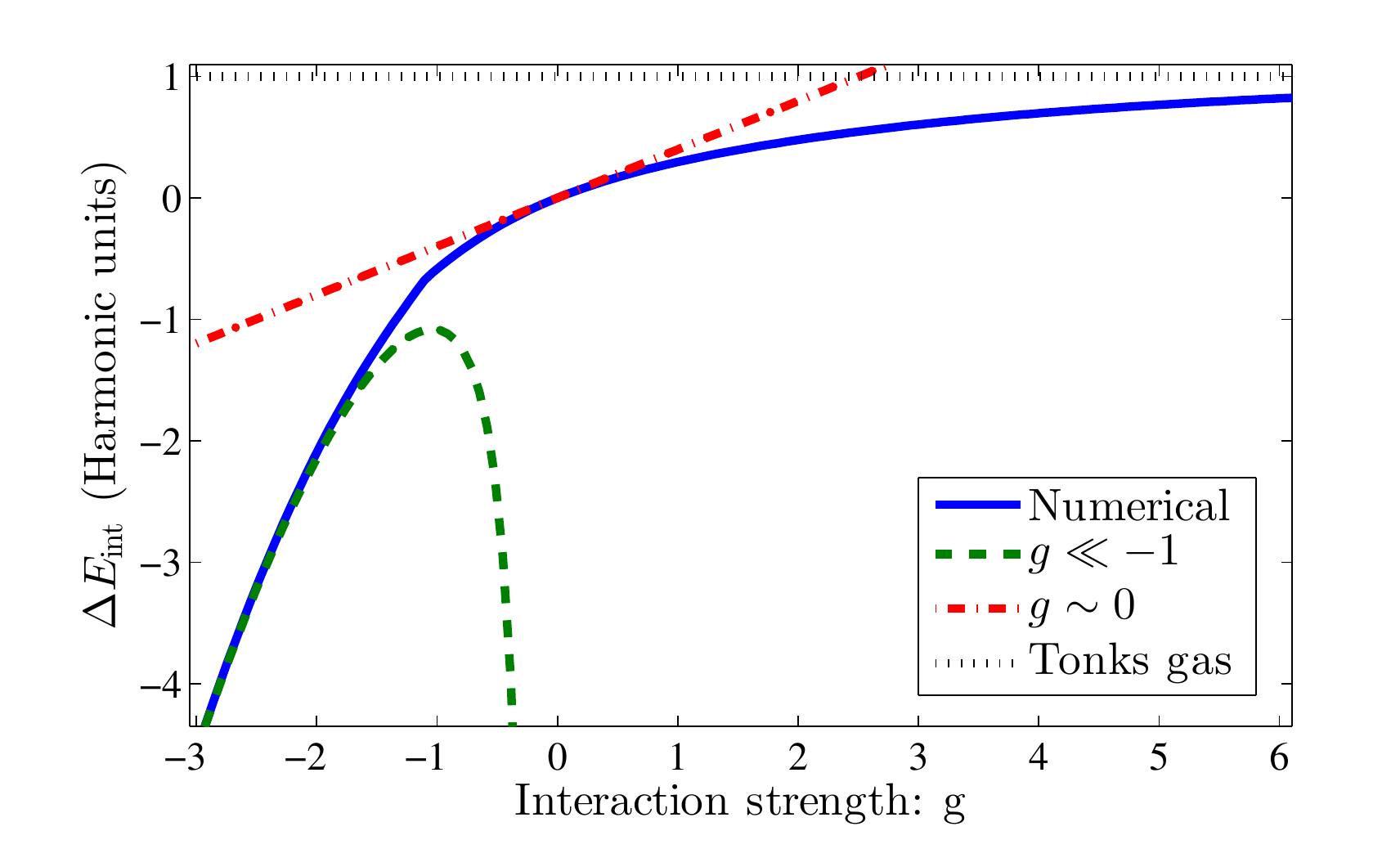}
\caption{(Color online): Energy difference $\Delta E_{\rm int}=E_{3,1}-E_{2,2}$ between two+two and three+one atom ground state clusters, as a function of interaction strength in harmonic units.  Analytic estimates from \refeq{eq:deltaEassym} are shown for comparison with the Tonks gas being the $g \to \infty$ limit.}
\label{fig:energy_difference}
\end{center}
\end{figure}

A feature which distinguishes this effect from intra-cluster excitations is the energy difference between the two configurations, denoted $\Delta E_{\rm int} =E_{3,1} - E_{2,2}$. For $g<0$ the ground state of a three atom relative Hamiltonian (that part of the Hamiltonian independent of the center of mass)  plus a single free atom is lower in energy than two sets of two atoms in their relative ground states.  The opposite is true for $g>0$, but the energy difference can only be of the order of the harmonic oscillator energy spacings, and so suppression is unlikely unless $x_0$ is small. The energy difference $\Delta E_{\rm int}$ can take a variety of values when intra-cluster states are excited, but in the interest of studying transfer interactions, we look at the energy difference between two isolated ground states of $N=2$ atoms and an $N=3$ and $N=1$ atom ground states. This can be estimated analytically in three limits:
\begin{align}
  \Delta E_{\rm int} \sim \begin{cases}
g/\sqrt{2 \pi} &\mbox{if } |g| \ll 1, \\
1 & \mbox{if } g \gg 1, \\
-g^2/2 - 7/12 g^2 & \mbox{if } g \ll -1;  \end{cases}
\label{eq:deltaEassym}
\end{align}
the approximations used being overlapping non-interacting ground states, effective fermionization~\cite{Girardeau2001}  (Tonks gas) and bound state clusters~\cite{McGuire1964} with the first order energy correction from the trapping potential~\cite{HoldawayWeiss2012}, respectively.  Numerically determined values of $\Delta E_{\rm int}$ are shown in \reffig{fig:energy_difference}; this energy proves to be an important quantity in the next section (note that this does not include the energy from the momentum/displacement of the clusters).  
Viewed classically, this transfer interaction causes transfer to a state where the kinetic energy of the clusters was different from the original by an amount equal to $\Delta E_{\rm int}$, in order to conserve energy.

\subsection{Mixing between different number configurations via time-dependent perturbation theory \label{sec:singlet_trimer_mix}}

We now investigate the atom transfer effect outlined in section \ref{subsubsec:transfer}, predicted to be most significant for $g<0$.  We can write our wavefunction at any point in time as
\begin{equation}
\ket{\psi(t)} = c_{2,2}(t) \ket{\psi_{2,2}(t)} + c_{1,3}(t) \ket{\psi_{1,3}(t)} + c_{0,4}(t) \ket{\psi_{0,4}(t)} \;,
\end{equation}
with $\ket{\psi_{n,N-n}(t)}$ normalized wavefunctions which are superpositions of states with $n$ and $N-n$ atoms to the left 
and vice versa, and $\{c_{n,N-n}\}$
a set of complex constants, the modulus squares of which are the probabilities to find $n$ or $N-n$ atoms either side.  In order to qualitatively predict the incremental changes to 
$\{c_{n,N-n}(t)\}$
from before to after a collision, we use time dependent perturbation theory, assuming $|g\lesssim  1$ is a small parameter and neglecting any contribution from $c_{0,4}(t)$ (specifically at the time of collisions).  We further assume the center-of-mass motion of each $n,N-n$ atom cluster in $\ket{\psi_{3,1}(t)}$ undergoes harmonic oscillation and is periodic in time with period $T=\pi$, and that any internal relative excitations in both $\ket{\psi_{n,N-n}(t)}$ are small compared to the ground state.  This approximation is expected to work better for $g<0$, for reasons outlined in Sec. \ref{subsubsec:intra-cluster}, and at short-times.   
As we initially have only $c_{2,2}\neq 0$, we assume $|c_{3,1}(t)| \ll |c_{2,2}(t)|$ as a regime of validity. 

Formally, we perturb $(\hat{H}_L + \hat{H}_R)$ by $\hat{H}_{I}$ [see \refeq{eq:split_ham}].  
Our wavefunction
\begin{align} 
 \ket{\psi(t)} \simeq c_{2,2}(t) \ket{\psi_{2,2}(t)} + c_{1,3} \ket{\psi_{1,3}(t)} \; ,
\end{align} 
must solve
\begin{align}
i \frac{d}{dt} |\psi(t)\rangle = 
[(\hat{H}_{L} + \hat{H}_{R}) + \hat{H}_{I}]
|\psi(t)\rangle \;.
\end{align}
We assume the difference between the time-derivative of $|\psi_{2,2}(t)\rangle$ and $(\hat{H}_{L} + \hat{H}_{R}) |\psi_{2,2}(t)\rangle$ is small (which assumes there is only a small amount of relative excitation), and neglect the time-derivative of $|\psi_{3,1}(t)\rangle$; by our initial assumptions, the prefactor $c_{3,1}(t) $ is small. 
The time-derivatives of $\{c_{n,N-n}(t)\}$ are thus given by:
\begin{align}
i[ \dot{c}_{2,2}(t)|\psi_{2,2}(t)\rangle + \dot{c}_{3,1}(t)  |\psi_{3,1}(t)\rangle]
\simeq \hat{H}_{I}c_{2,2}(t)|\psi_{2,2}(t)\rangle.
\end{align}
Hence, [using that $|\psi_{3,1}(t)\rangle$ and $|\psi_{2,2}(t)\rangle$ are orthogonal]
\begin{align}
i \dot{c}_{2,2}(t) &\simeq c_{2,2}(t) \langle\psi_{2,2}(t)|\hat{H}_{I}|\psi_{2,2}(t)\rangle, \\
i \dot{c}_{3,1}(t) &\simeq c_{2,2}(t)\langle\psi_{3,1}(t)|\hat{H}_{I}|\psi_{2,2}(t)\rangle.
\end{align}
Within first order perturbation theory, $\langle\psi_{2,2}(t)|\hat{H}_{I}|\psi_{2,2}(t)\rangle$ is periodic with a periodicity $(T=\pi)$ half that of the oscillator period. The matrix element $\langle\psi_{3,1}(t)|\hat{H}_{I}|\psi_{2,2}(t)\rangle$ is a product of a function with period $T=\pi$, and the complex exponential $\exp(-i\Delta E_{\rm int} t)$ of the energy difference between the intra-cluster degrees of freedom in both configurations (as plotted in \reffig{fig:energy_difference}). 

Denoting the periodic component of the interaction terms $\langle\psi_{n,N-n}(t)|\hat{H}_{I}|\psi_{2,2}(t)\rangle$  as $f_{n,N-n}(t)$, we must therefore solve
\begin{align}
i \dot{c}_{2,2}(t) &\simeq c_{2,2}(t) g f_{2,2}(t) \;, \\
i \dot{c}_{3,1}(t) &\simeq c_{2,2}(t) g f_{3,1}(t) \exp\left( -i \Delta E_{\rm int} t \right) \;,
\label{eq:time_deriv_coefficients}
\end{align}
with the boundary condition $c_{2,2}(0) =1$.  We first assume that the initial separation $x_0$, and the interaction strength $|g|$, are not large.  Within this regime we assume we can approximate $f(t)$ by a first order Fourier series $f(t) \approx 1 - \cos(2t)$, which implies all $f_{n,N-n}(t)$ differ only by a constant value; hence $f_{n,N-n}(t)=Af(t)$, $f_{1,3}(t)=Bf(t)$, with $A$ and $B$ dependent, in principle on $g$, and quite heavily on $x_0$.   We can use this to solve \refeq{eq:time_deriv_coefficients}:
\begin{equation}
\begin{split}
 c_{2,2}(t) \simeq & \exp\left(i\int_0^t  \mathop{d t'} A g  f( t' )\right)
	     \\
	     \simeq   & \exp(i A g [t - \sin(2t)/2 + \ldots] ),
	     \end{split}
\end{equation}
and if we neglect $\Delta E_{\rm int}$ under the assumption that the relative energy on both sides is similar, 
\begin{align}
 c_{3,1}(t) \simeq \frac{B}{A}  \left[
 \exp\left(i g A\int_0^t d t'  f( t' )\right)- 1
 \right] 
 \;.
\end{align}

For short times, we can expand $c_{31}(t) \approx B [ i g t + {\cal O}(g^2t^2,g\cos(2t)) ]$, i.e., proportional to $gt$ and oscillatory terms and hence giving a linear increase when $t=n\pi$.  At longer times the phase evolution of  $c_{2,2}(t)$ becomes important, leading to cancellation in the terms of $c_{3,1}(t)$ and giving oscillatory behavior with a period dependent on $g$.  The linear increase with $g$ after a collision is not expected to continue when $g \gtrapprox 1$ as higher-order terms become increasingly important and the perturbation theory breaks down.

We have so far neglected the difference in internal energy. This will introduce an additional phase between $c_{3,1}(t)$ and $c_{2,2}(t)$.  With this included we have
\begin{equation}
\begin{split}
 c_{3,1}(t) &\simeq \int_0^t \mathop{d t'}  \left[i g B  c_{2,2}( t' ) f(t') \exp\left( -i \Delta E_{\rm int} t' \right) \right]  \\
&\simeq i g B \int_0^t \mathop{d t'} f(t') \exp(i [A g-\Delta E_{\rm int}  t'  - {\rm osc}]) \;,
\end{split}
\end{equation}
with ``osc'' denoting oscillatory terms such as $k\cos(2t)$, which are periodic with $t \to t + \pi$ or shorter fractions of $\pi$ for the higher-order terms.  
Summing together terms of different phases will produce cancellation, hence if the $\exp(i [A g-\Delta E_{\rm int}  t] )$ term has the same periodicity as $f(t)$ and the ``osc'' terms, both $\pi$, the overall increase will be linear in time with no higher-order polynomial terms.  
This could therefore lead to resonant (suppressed) transfer if $Ag-\Delta E_{\rm int} \approx n$  with $n$ even (odd), and slightly suppressed transfer if $n$ is a rational number not close to an even integer, e.g.\ $1/2, 1/3,3/2$.  As noted earlier, the $g \to \infty$ limit gives $\Delta E_{\rm int} \sim 1$ and thus should lead to suppressed transfer if $|Ag|\lesssim  1/2$.  We note that when $|g| \sim 0$ this resonance condition appears to be matched up to a factor $ g[A-(2 \pi)^{-2}]$, giving very long cancellation periods, however, as we see on \reffig{fig:DeltaNfirstcol} (and by the fact the perturbation strength scales $\propto g$) the rate of atom transfer scales proportional to $g$ and so cancellation can still occur before a significant population transfer is achieved.    

This simple analysis neglects higher-order effects such as  pseudo-periodicity, and intra-cluster excited states are not treated explicitly.  However, qualitatively we expect an initially weak linear increase with long time oscillation effects for small $|g|$, and for $g \gtrapprox 1$ the timescale of these oscillations should drop.

\subsection{Amplitude bound to oscillations}

One can look at each left/right number eigenstate [\refeq{eq:resregdef}] separately, assuming we have a probability of $p$ for $\ket{2,2}$, and of  $(1-p)/2$ for $\ket{3,1}$ (with the same for the $\ket{1,3}$ state), no occupation of $\ket{4,0}$ or $\ket{0,4}$, and that there is no overlap between the states and no mixing via the Hamiltonian.  We can then state the energy $E_{1,3} = \bra{1,3} \hat{H} \ket{1,3}$ as follows
\begin{equation}
 E_{1,3} = E_{\rm pot,1} + E_{\rm pot,3} + E_{\rm kin,1}+ E_{\rm kin,3}+ E_{\rm int,3} \;.
\end{equation}
Each term in this equation refers to the kinetic, potential and interaction energy of each side, with one or three atoms, respectively (note there is no interaction energy for the single atom side, taken without loss of generality as being left).  Noting that the kinetic and potential energy terms must be positive, we can derive the inequality
\begin{equation}
 E_{\rm pot,1} \le E_{1,3} - (E_{\rm kin,3}+E_{\rm int,3}) \;.
\label{eq:boundpot1}
\end{equation}
Using the conservation of $E = \ev{\hat{H}}$ and $\Delta E^2 =\ev{\hat{H}^2}-E^2$, it can be shown that (see Appendix \ref{app:energy_bound} )
\begin{equation}
|E_{3,1} - E | \le \sqrt{\frac{p}{1-p}} \Delta E \;,
\label{eq:indv_state_bounds}
\end{equation}
which is equivalent to
\begin{equation}
E - \sqrt{\frac{p}{1-p}} \Delta E \le E_{3,1} \le E+ \sqrt{\frac{p}{1-p}} \Delta E \;.
\end{equation}
Combining the upper bound of the above equation with \refeq{eq:boundpot1}, we obtain
\begin{equation}
 E_{\rm pot,1} \le  \left(E+ \sqrt{\frac{p}{1-p}} \Delta E\right)- (E_{\rm kin,3}+E_{\rm int,3}) \;.
\label{eq:boundpot2}
\end{equation}
Finally, noting that $E_{\rm pot,1} = \ev{x^2}_{1}/2 \ge \ev{x}_{1}^2/2$, with the $\ev{\hat{O}}_1$ meaning the expectation value of the 1 particle side of the wavefunction, we can obtain an inequality for the 1 atom position expectation value 
\begin{equation}
 \ev{x}_{1} \le \sqrt{2}\sqrt{\left(E+ \sqrt{\frac{p}{1-p}} \Delta E\right)- E_{\rm int,3}} \;.
\label{eq:boundpot3}
\end{equation}
We can see that larger, positive $g$ will constrain this bound, up to a point of saturation at the Tonks-gas limit, whereas potentially it is unbounded as $g \to -\infty$ (energies in this regime scale proportional to $-g^2$~\cite{McGuire1964}) as the atoms gain a large amount of energy.

\section{Possible experimental realization of the four atom system \label{sec:experimental}}

\subsection{Optical lattice scheme}

\begin{figure}
\begin{center}
\includegraphics[width=\linewidth]{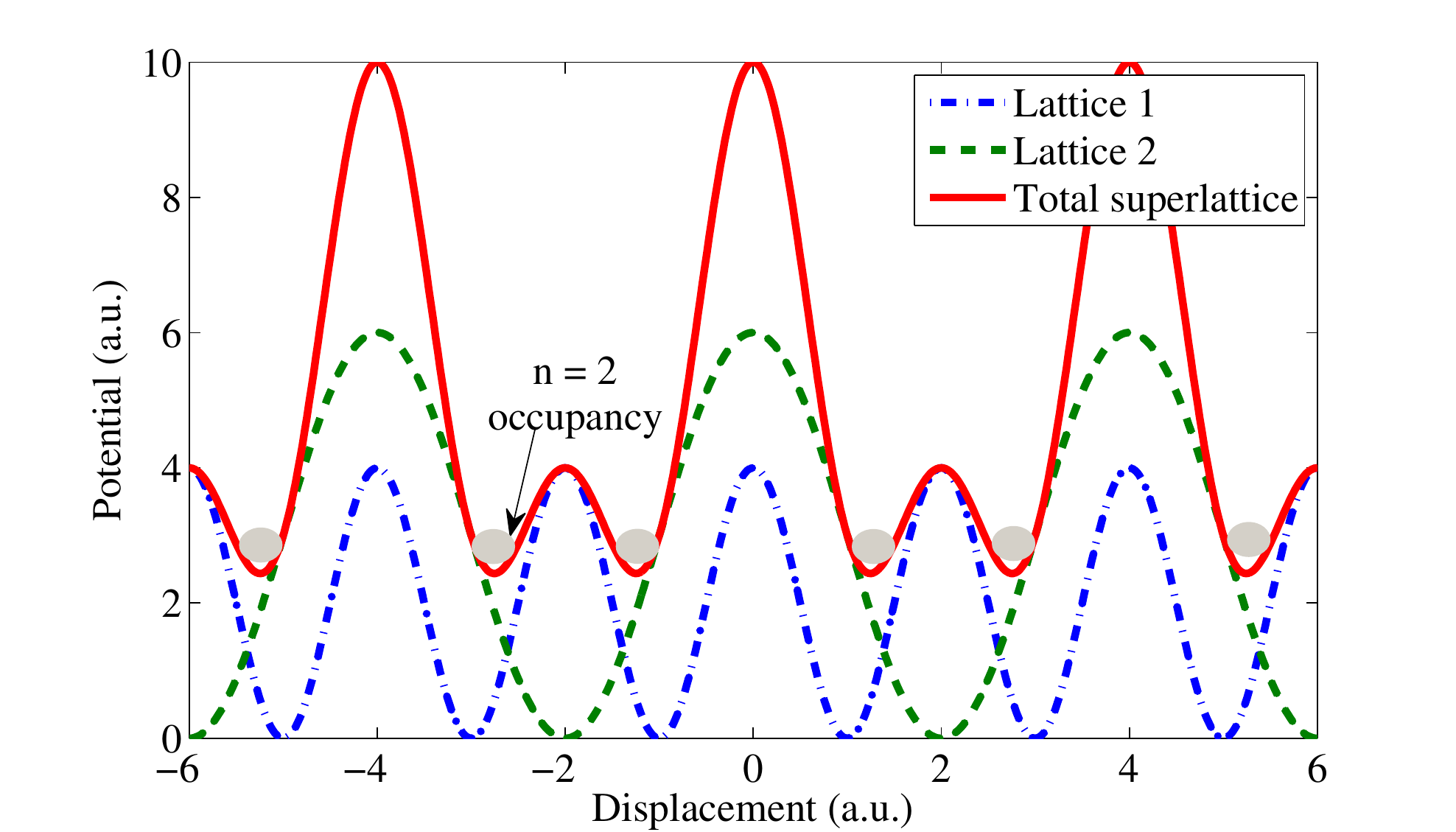}
\caption{(Color online): Potential from an optical super-lattice created by overlapping two lattices (all units are arbitrary).  Grey circles represent a loading of two atoms in the ground state of each well.  Our suggested scheme tunes the interactions to the desired value and then turns off the double frequency (dotted line) lattice leaving only the broader lattice (dot-dashed line), after which the atomic dimers collide.}
\label{fig:optical_super_lattice}
\end{center}
\end{figure}  

Our results could be tested by creating an optical super-lattice~\cite{Sebby-Strabley2006}, of two overlapping lattices, with one double the frequency of the other, then loading this with two atoms per site (in the ground state) in a Mott insulator regime~\cite{Bloch2012}.  This is shown schematically in \reffig{fig:optical_super_lattice}. The interactions could then be tuned to be attractive via a magnetic Feshbach resonance, at such a rate that tunneling between sites is small, but the two atoms on each site tend to the ground state given by \refeq{eq:twobodygs}.  The double-frequency lattice could then be ramped down, leaving only the wider lattice, thus creating the initial conditions of two equally separated dimers in an approximately harmonic potential.

Some freedom with $x_0$ could be achieved by modifying the relative strength of the double-frequency lattice compared with the primary lattice. Reducing it will push the minimum closer together, but also make tunneling between the sites more significant.  Careful ramping-down schemes of the laser power of the double-frequency lattice could also be incorporated, which would give further freedom to move the sites closer together after creating the dimers. Slower ramping will also make things closer to adiabatic,  thus reducing the excitation in each dimer created by the switch-off.  The relative velocity between the two dimers in terms of the final harmonic oscillator units will equate to an effective initial separation --- approximately the separation the dimers will reach after the first collision.  A faster (slower) ramping scheme would give a larger (smaller) effective $x_0$, however, to be most applicable with the results of this paper, a slow scheme would be ideal to minimize excitations and minimize the degree of anharmonicity in the potential that the dimers sample.  

After some free-evolution time, the double frequency lattice could then be quickly restored with an extremely high lattice depth, separating the left and right components of the wavefunction, with no further tunneling possible.  This would allow for a direct measurement of $\hat{N}_R$ as defined in \refeq{eq:numlr}, by then imaging the lattice with resonant light; light-induced collisions~\cite{Sherson2010} will reduce this to a parity measurement with an empty site being either a zero or two population, and a single atom being a one or three population.  This is actually sufficient information, assuming we know the total atom number in the two sites was exactly 4. In terms of the states given in \refeq{eq:resregdef}: no atoms on either site is a measurement of a $\ket{2,2}$ configuration (or a $\ket{4,0} /\ket{0,4}$ configuration, but this is only significant during collisions), both sites occupied is a measurement of a $\ket{3,1} /\ket{1,3}$ configuration, a single occupied site and an empty site would imply some inelastic process has occurred (such as three-body recombination or background gas collisions) and such a result would thus be null. 

If the effective $x_0$ were an appreciable fraction of the lattice width, this scheme could also show some more interesting physics beyond the scope of this paper, with collisions coupling energy into the center-of-mass mode and the tunneling of the single atom in the single-trimer states (considered in Sec. \ref{sec:singlet_trimer_mix}) to adjacent lattice sites. It could even have a kinetic energy greater than the maximum barrier height between sites and join an effective conduction band~\cite{Sherson2012}, allowing for entanglement between lattice sites.  These effects may also be worthy of experimental investigation.

\subsection{Experimental parameters}

In terms of typical experimental parameters, the $s$-wave scattering lengths would need to be very substantial in order to give measurable effects. Strong interactions generally require tuning scattering lengths near to Feshbach resonances, and in such strongly interacting regimes confinement effects can shift the effective 1D scattering length if $a_s / a_{\bot}$ is not small~\cite{Olshanii1998}.  The chosen Feshbach resonance would ideally be broad, minimizing uncertainty in the effective interaction associated with a lack of precise control of magnetic field fluctuations.

Alternatively, some atoms such as Cesium can have large ``background'' scattering lengths far from resonances~\cite{ChinVuletic2004}, e.g., $a_s \sim \pm 3000 a_0$ where $a_0 \approx 5.3 \times 10^{-11} m$ is the Bohr radius.  In terms of a rescaled $g$ parameter in harmonic oscillator units, if one had $\omega_x \sim 2\pi \times 1$Hz and very strong radial confinement $\omega_{\bot} \sim 2\pi \times 0.4 $kHz, we have
\begin{equation}
 g = 2\omega_{\rm \bot}  a_s \sqrt{\frac{m}{\hbar\omega_x}} \sim \pm 1.2 \;,
\end{equation} 
which is of unitary order.  

We essentially have three experimentally tunable parameters, $a_s$, $\omega_x$ and $\omega_{\bot}$ which can be varied smoothly with small adjustments to a magnetic field or modifying laser powers , focusing, or detunings.  However, dropping $\omega_x$ is undesirable as it increases experimental timescales, and increases the likelihood of background gas collisions; additionally, unwanted three-body recombination effects scale $\propto |a_s|^4$ 
(generally being worse for $a_s <0$) meaning one would need to determine an appropriate  compromise solution.

\section{Numerical method \label{sec:numeric_meth}}

\subsection{Basis set expansion}
To perform many-body computations we expand the field operator over the set of Hermite functions of a given width $W$
\begin{equation}
\varphi_k(Wx) = \sqrt{\frac{W}{k!2^k\pi^{1/2}}} H_k(W x) \exp\left(-W^2 x^2/2\right) \; , 
\label{eq:Wbasis}
\end{equation}
with $H_k(x)$ the Hermite polynomials, and diagonalize the Hamiltonian in a Fock state basis $\ket{n_0,\ldots,n_{\infty}}$, truncated via the condition $\sum_k k n_k \le \eta$.  Such a calculation would require an unfeasible amount of states to converge, were it not for the fact that the center-of-mass part of the Hamiltonian commutes with the rest of it.  This means we can just consider a subset of this truncated Fock space where the center of mass of the gas is in the same state. This does not have to be the ground state, as we simply ignore the center-of-mass time evolution and can account for it later.  The procedure essentially involves diagonalizing the finite basis in terms of the operator
\begin{equation}
 \hat{A}^{\dagger} \hat{A} = \sum_{k,j} \sqrt{(k+1)(j+1)} \hat{a}_{k+1}^{\dagger} \hat{a}_k \hat{a}_j^{\dagger} \hat{a}_{j+1} \; , 
\end{equation}
(where $\hat{A}^{\dagger} = \sum_{k} \sqrt{k+1}\hat{a}_{k+1}^{\dagger} \hat{a}_{k}$ is the creation operator for a dipole mode of width $W$) and taking the eigenvectors with eigenvalue zero; this procedure is discussed in detail in~\cite{HoldawayWeiss2012}.  For the calculations in this paper we use the eigenstate width $W = 1$ as the harmonic oscillator length is always a relevant scale.

\subsection{Convergence testing}

We first need to represent our initial condition in terms of this basis set, noting that due to the truncation the state cannot be represented exactly, with larger initial displacements and larger interaction strengths harder to represent in this basis. We require a reasonable fidelity of our numerical initial condition to the true state, achieving fidelities of $> 99.5\%$ for all the numerics used in this paper.

\begin{figure}
\begin{center}
\includegraphics[width=\linewidth]{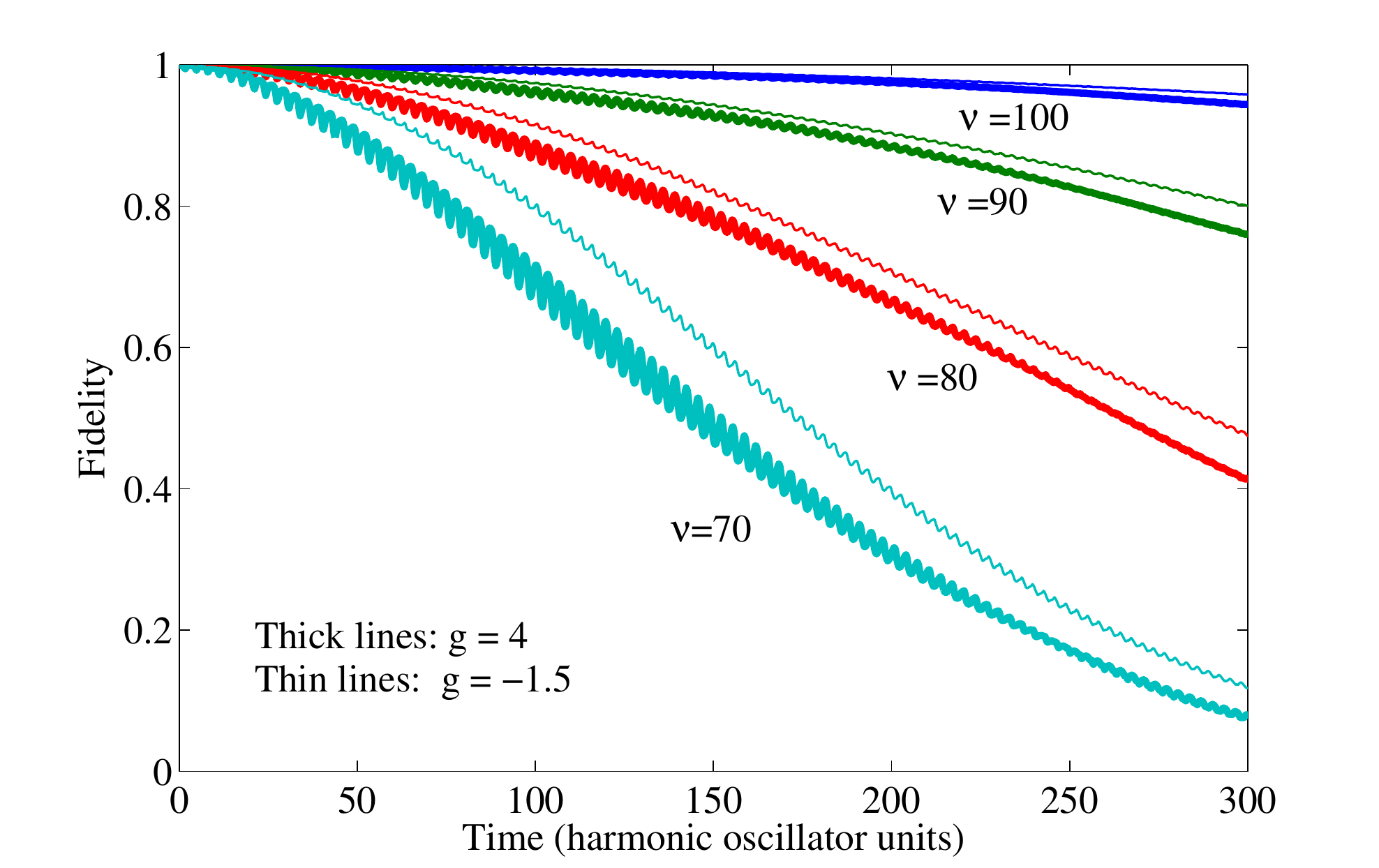}
\caption{(Color online): Plot of $|\bket{\psi(\nu,t)}{\psi(\nu_{\rm max},t)}|^2$; the fidelity of the wavefunction computed with smaller basis (energy cut off at $\nu$) to the wavefunction computed using a larger basis truncated at $\nu_{\rm max}=113$.  This shows the extreme values of $g$ employed in the numerics, lower absolute values of $g$ converge more rapidly.  }
\label{fig:convergence}
\end{center}
\end{figure}

Measuring convergence during time evolution with such a method is more difficult. Performing the calculations with a variety of basis sizes and calculating the fidelity over time can give an indication for how long the calculations are reliable, for which we plot, in \reffig{fig:convergence}, our most extreme values of $g$.  This is probably the strictest measure of convergence applicable, given the large number of degrees of freedom in a many body wavefunction, for example a product state with a large number of atoms would have a fidelity exponentially tending to zero for any finite difference in the product wavefunction. 

\section{Numerical results \label{sec:results}}

\subsection{Preamble}
All the results graphed here are calculated for $N=4$ and $x_0=3$ in order to investigate the effects of varying interaction strength for small numbers.  In general smaller $x_0$ greatly increases interaction times between clusters and thus rates of atom transfer. It also reduces the amount of free energy in the system, however a greater amount of the wavefunction will be found towards the center at all times and thus expectation values of $\hat{N}_R$ will be harder to interpret.  The results here are broken down into three sections, the first examines the variation in left right number, the second examines the variance is position about one side and the final section examines the single body von Neumann entropy.

\subsection{Left and right particle number dynamics \label{sec:results_numlr}}

\begin{figure}[h!]
\begin{center}
\includegraphics[width=\linewidth]{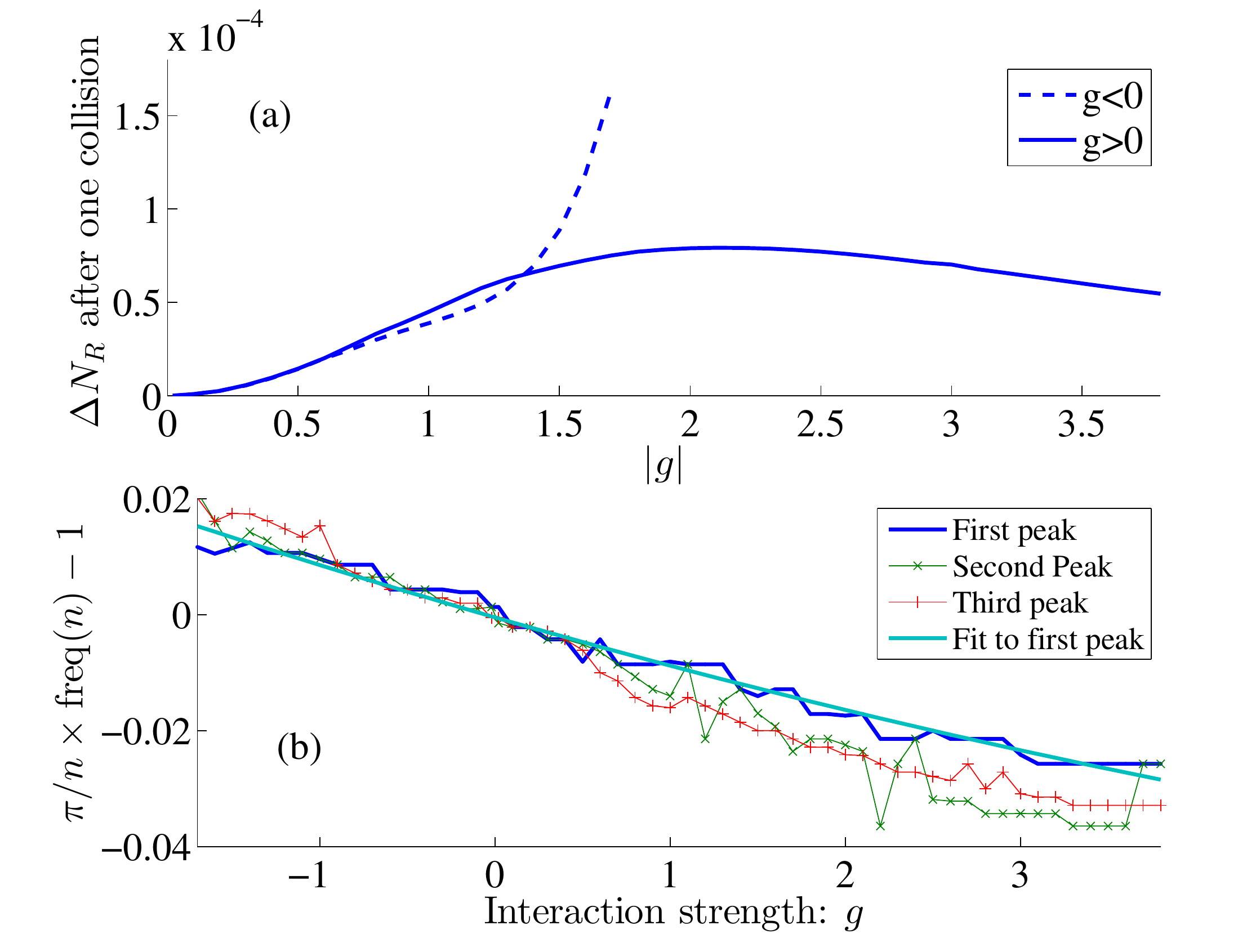}
\caption{(Color online): (a) Minimum value taken by $\Delta N_R$ [\refeq{eq:DeltaNR}], after one collision (b) Frequency difference of peaks in the Fourier transform of $\Delta N_R$ from the non-interacting values $(t=n \pi)$ divided by $n$. (a) shows that for $g>0$, the increase to number uncertainty is greatest for $g \approx 2.3$ and decreases when interaction strength is increased further.  The $g<0$ behavior is initially similar but deviates at around $|g| =0.6$; rather than saturating it appears to increase even more rapidly with $|g|$.  It is not clear what will happen for $g < 0$ and $|g| \gg 1$, which will be a topic for further investigation.  (b) demonstrates the existence of pseudo-periodicity in the system (in addition to low frequency components relating to the long time behavior).  The non-interacting system has frequency peaks at $f_n = n/\pi$, the quadratic fit (solid line) indicates these peaks shift by an amount roughly equal to $-n g /100 \pi$.}
\label{fig:DeltaNfirstcol}
\end{center}
\end{figure}

\begin{figure}[h!]
\begin{center}
\includegraphics[width=\linewidth]{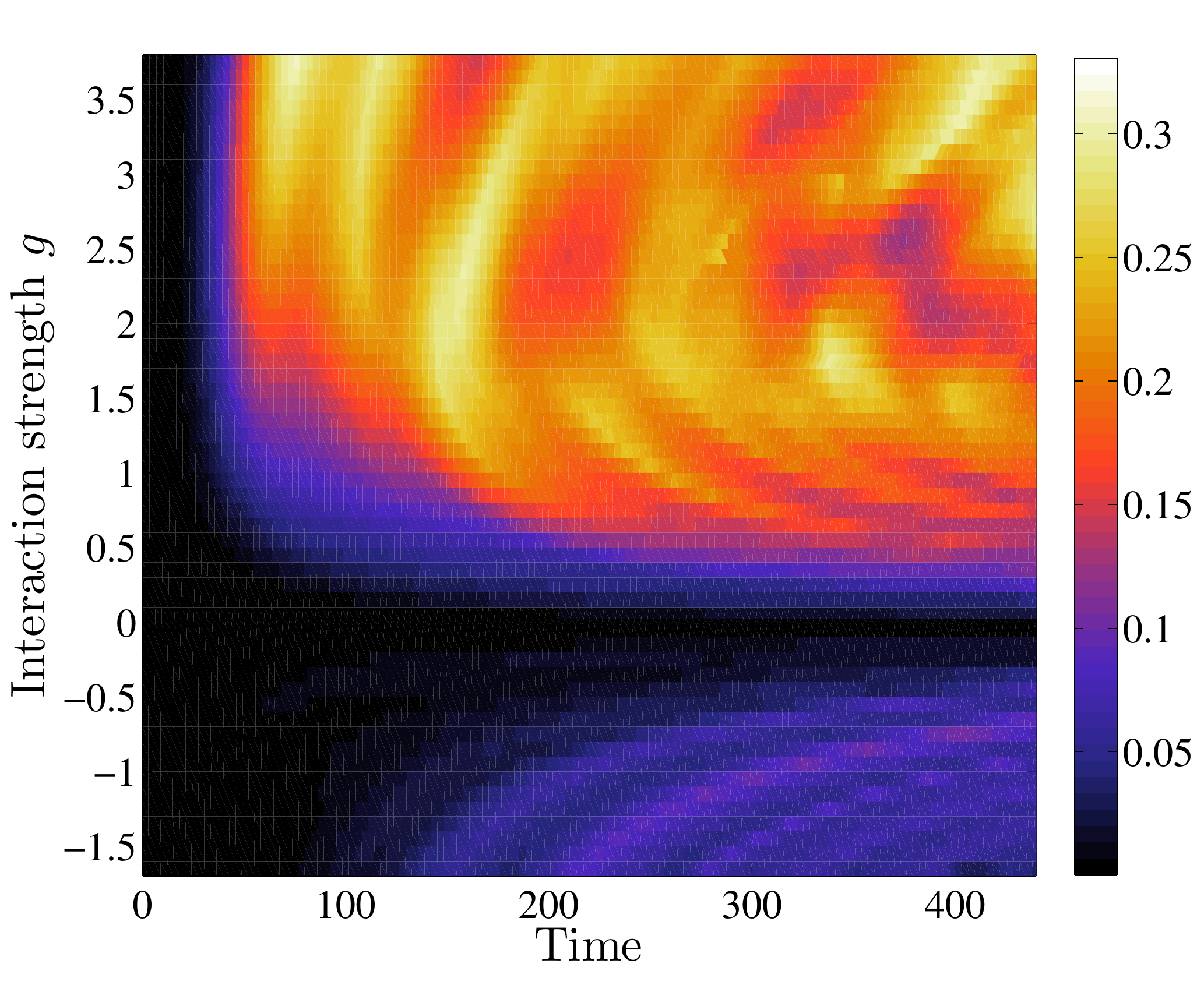}
\caption{(Color online): Minimum value obtained by $\Delta N_R$, as given by \refeq{eq:DeltaNR}, after a given collision.  For weak interactions $(|g| < 0.1)$ the behavior is the same for attractive and repulsive, but for slightly larger values there is a clear difference in the timescales, with repulsive interactions producing larger number uncertainties more quickly, despite the fact that \reffig{fig:DeltaNfirstcol} shows there is little difference in $\Delta N_{R}$ after one collision.  This difference is likely due to the increased (decreased) energy spacing between the ground and first excited state of the two atom system with attractive (repulsive) interactions, discussed in Section \ref{subsubsec:intra-cluster}, and the energy difference between the two-two and three-one number configurations, as discussed in Section \ref{sec:singlet_trimer_mix}, which leads to a phase mismatch.  For large  repulsive values $(g>2)$, $\Delta N_R$ reaches a maximum value and then undergoes complex partial revivals on timescales of ~30 time units, (tens of collisions).  }
\label{fig:DeltaNcolormap}
\end{center}
\end{figure}

Because our initial condition has a definite number of two atoms either side of the trap, the left-right number uncertainty, $\Delta N_R$, in our system is initially very near zero.  We note that a mean-field-like state or a symmetric superposition of 3 and 1 atoms either side both give $\Delta N_R =1$, which is also the value this quantity will take in our non-interacting system when each of the clusters collide.  We therefore first consider the minimum to minimum values taken by $\Delta N_R$ before and after each collision. The change after the first collision is given in \reffig{fig:DeltaNfirstcol} and the change over the first 150 collisions is plotted in \reffig{fig:DeltaNcolormap}.  Despite the fact that the increase after the first collision is similar for both attractive and repulsive interactions of similar magnitude, the long time change is very different, with the timescales being much longer in the attractive case.  

In either case, the left-right number does not reach an equilibrium on the timescales considered, with oscillations and revivals present. The time-dependent perturbation theory of Section \ref{sec:singlet_trimer_mix} indicates that atom transfer processes are suppressed by an internal energy difference between the $\ket{2,2}$ and $\ket{3,1}$ configurations of the wavefunction, which leads to destructive mixing over a few collisions, unless a phase matching condition occurs.  If intra-cluster excited states (discussed in Section \ref{subsubsec:intra-cluster}) are present, the energy difference between each configuration, $\Delta E_{\rm rel}$, may be small (along with $A g$) meaning cancellation occurs on longer timescales, leading to fluctuations in $\Delta N_R$ over 10s of harmonic oscillator periods.   

\begin{figure}[h!] 
\begin{center}
\includegraphics[width=\linewidth]{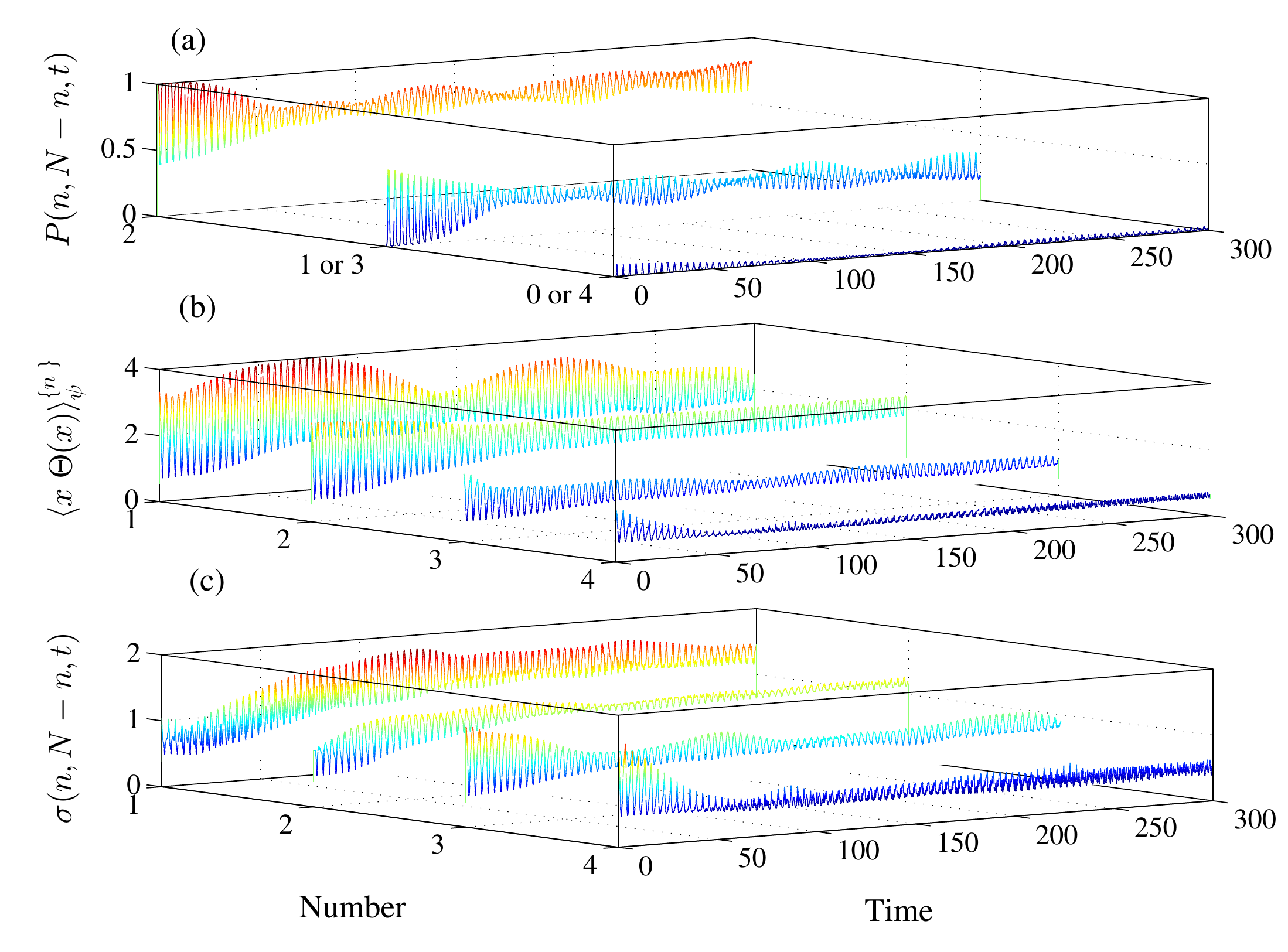}
\caption{(Color online): For $g = 3$, $x_0 = 3$, (a) time evolution of probability of finding $n$ (or $N-n$) atoms to the right with the amplitudes of wavefunction components decomposed into eigenfunctions of number L/R number operator.  (b) Expectation value of position to the right on sections of wavefunction decomposed into eigenfunctions of L/R number operator. (c) Variance in position to the right as defined in \refeq{eq:sigma_LR}, paralleling (b). The expectation value to the right [(b)] effectively tracks the particle-like motion, but after long times the motion appears effectively damped.  (c) can quantify this effect --- the peaks of $\sigma(n,N-n)$ increase from their initial value and continue to oscillate about a maximum, except for $\sigma(4,0)$ (which is only significantly probable during collisions) indicating a transfer of energy to the degrees of freedom described in Sections \ref{subsubsec:inter-cluster} and \ref{subsubsec:intra-cluster}.  This remains true even at very long times $t \sim 1000$, with progressively smaller partial revivals and so can be said to have equilibrated.}
\label{fig:expdisp}
\end{center}
\end{figure}

\begin{figure}[h!] 
\begin{center}
\includegraphics[width=\linewidth]{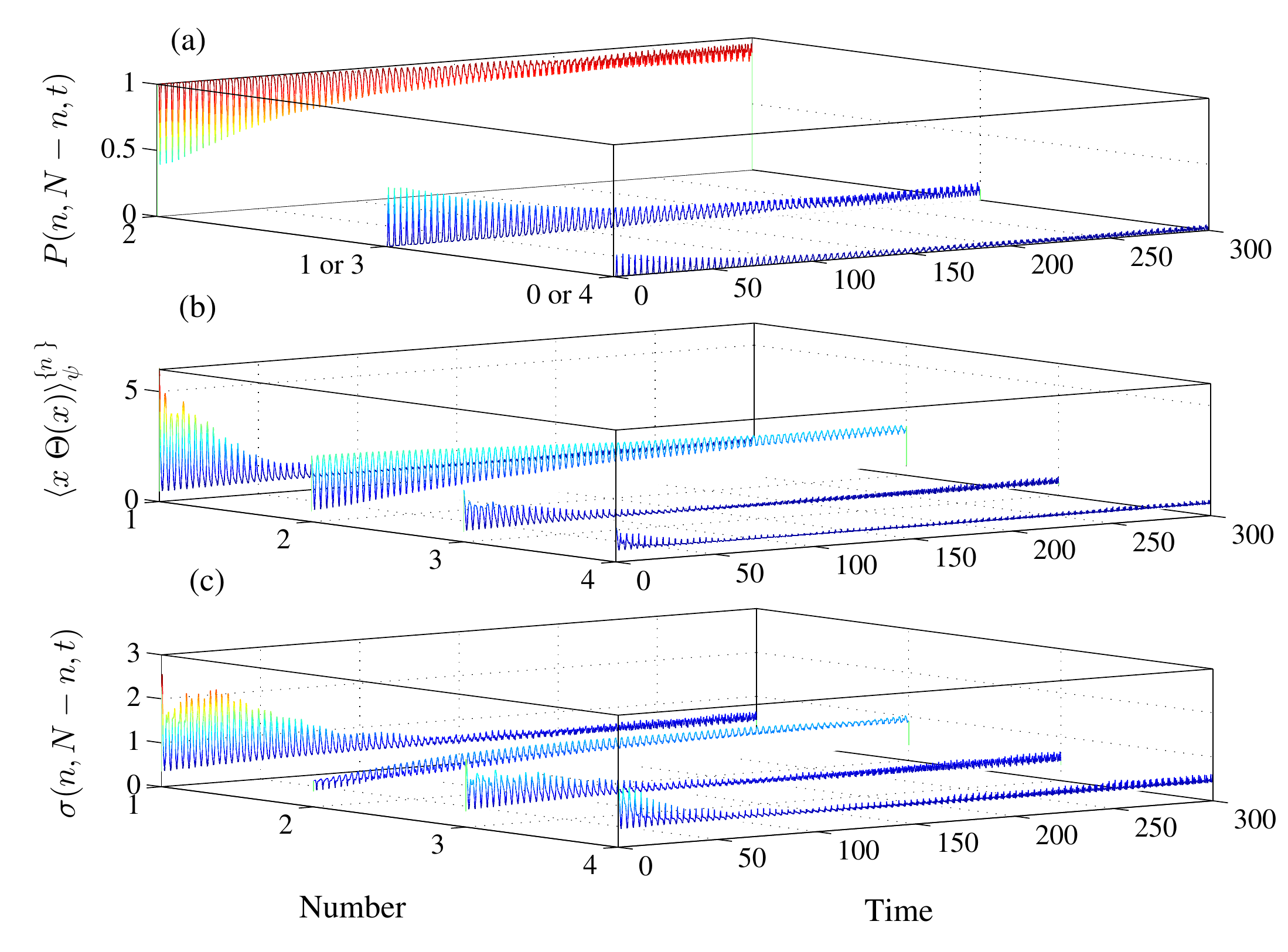}
\caption{(Color online): The same quantities as \reffig{fig:expdisp} but with $g = -1.7$.  The short-term behavior of the expected one-atom position (b) is similar to the repulsive case, but is increased in magnitude.  At long times, the right-position expectation values drop to an approximately constant value for all but $n=2$, this being the value of a Gaussian state in the center of the trap, for reasons explained in section \ref{sec:results_numlr}.  This is also the case in (c) --- essentially the only significant contribution to the $n \neq 2$ states comes from uncertainty in the separation of the atomic dimers, which smooths over transfer effects. }
\label{fig:expdisp_glt0}
\end{center}
\end{figure}

Figures \ref{fig:expdisp} and \ref{fig:expdisp_glt0} (a) show the amplitude of each number component in the wavefunction as it evolves in time for $g=3$ and $g=-1.7$; note \reffig{fig:DeltaNcolormap} takes only the minimum values of these curves to avoid the spikes on collisions.  The maximum amplitude of the $\ket{3,1}$ and $\ket{4,0}$ components (at least initially) occurs on collisions (corresponding to a minimum amplitude of $\ket{2,2}$).  Decreasing of this peak amplitude may be interpreted as the time of collisions between clusters becoming less well defined, due to the distance between their centers of mass becoming less well-defined (i.e., its corresponding probability density becomes broader) and the forming of intra-cluster excitations. 

At late times ($t > 100$) on figure \ref{fig:expdisp_glt0}, all the expectation values for $n \neq 2$ are almost the same as those for Gaussians centered on zero.  This is due to only the two-dimer (attractive $n=2$ ground states) setup being significant, as the exciting of "intra-cluster" excitations is suppressed by the large energy gap, and atom transfer interactions are suppressed by an energy difference, leading to a phase mismatch and hence a cancellation.  However, energy is still transferred to the relative position wavefunction (described in Section \ref{subsubsec:inter-cluster}), increasing the uncertainty in the separation of dimers, and so some component of the wavefunction is always undergoing a collision yielding a finite value for the left-right number uncertainty.  As a result of our scaling in \refeq{eq:restricted_ev}, the $n \ne 2$ values are just those of the dimer system in collision, and only a small contribution to $\ket{3,1}$ comes from states that are similar to a superposition of a cluster of 3 atoms to the left (right) and a free atom to the right (left). 

\subsection{Equilibration of energy into inter/intra-cluster excited states}
We wish to quantify the amount of energy transferred from the center-of-mass energy of each cluster to excitations between the atoms, as discussed in Section \ref{subsubsec:inter-cluster} and Section \ref{subsubsec:intra-cluster}.  We therefore investigate the standard deviation in the position to the right, for a given number of atoms to the right
\begin{multline}
\label{eq:sigma_LR}
 \sigma(n,N-n,t) = \\ 
\sqrt{\ev{\Theta(x) \hat{\Psi}^{\dagger}(x) x^2 \hat{\Psi}(x)}^{(n)}_{\psi} -\vert \ev{\Theta(x) \hat{\Psi}^{\dagger}(x)x \hat{\Psi}(x)}^{(n)}_{\psi} \vert^2 } \; ,
\end{multline}
essentially the width of the atomic density distribution on the right hand side, about the expected value for position, given that $n$ atoms are on the right-hand side.  
This is plotted in \reffig{fig:expdisp} (c). The repulsive case shows a consistent increase in the height of the peaks (excepting the $n=4$ peak), with only small periodic oscillations.  The attractive case however shows $\sigma(n,N-n,t)$ to be initially similar but then dropping to a minimum value for $n \neq 2$.  We note $\sigma(n,N-n,t)$ cuts off anything on the left side, and so is difficult to relate to the amount of excitation if the left and right states are separated by a distance smaller than the size of their internal structure, as they will contribute to all the $n \neq 2$ expectation values.   
Intra-cluster excitations as we have defined them are present if the wavefunction either side of the center does not look like a displaced $n$ atom ground state; it is possible such excitations could reduce the position uncertainty but are generally expected to make it broader and thus increase $\sigma(n,N-n,t)$.  These excitations are dominant processes in the increasing of $\sigma$ for the repulsive case plotted in \reffig{fig:expdisp} (c), and appear to persist at long times. 

For the $g<0$ case, at very early times, say $t < 20$, the contribution to $\sigma(3,1,t)$ from states in the single particle and cluster-of-3 configuration is visible.  By (approximate) momentum conservation the single atom must have considerably more energy after a collision than the 3-atom state, which explains the large $n=1$ position expectation values away from collision. However, in the strongly attractive case this transfer process is cyclic, and it never transfers large populations to these configurations.  As we noted before, contributions can come from an oscillating dimer state if the relative separation is small.  Initially this only occurs during collision, but inter-cluster excitations (which can be interpreted as an increased uncertainty in how much the centers of each cluster have shifted due to interactions), lead to an increase in relative position uncertainty\footnote{although \reffig{fig:DeltaNcolormap} indicates this process undergoes partial revivals}. Hence, at late times there is always significant wavefunction density in the trap center, that is to say at any time $t > t_{\rm late}$ some non-negligible part of the wavefunction is always undergoing collision.  Hence, if the contribution from the singlet-triplet state is too small to see we can conclude that the $\sigma(2,2,t)$ reaching a maximum corresponds to this mode reaching a steady configuration.  This is the dominant effect in the attractive case shown in \reffig{fig:expdisp_glt0}, but is also present for $g>0$.

\subsection{Relaxation to equilibrium}

\begin{figure}[h!]
 \centering
\includegraphics[width=7.5cm]{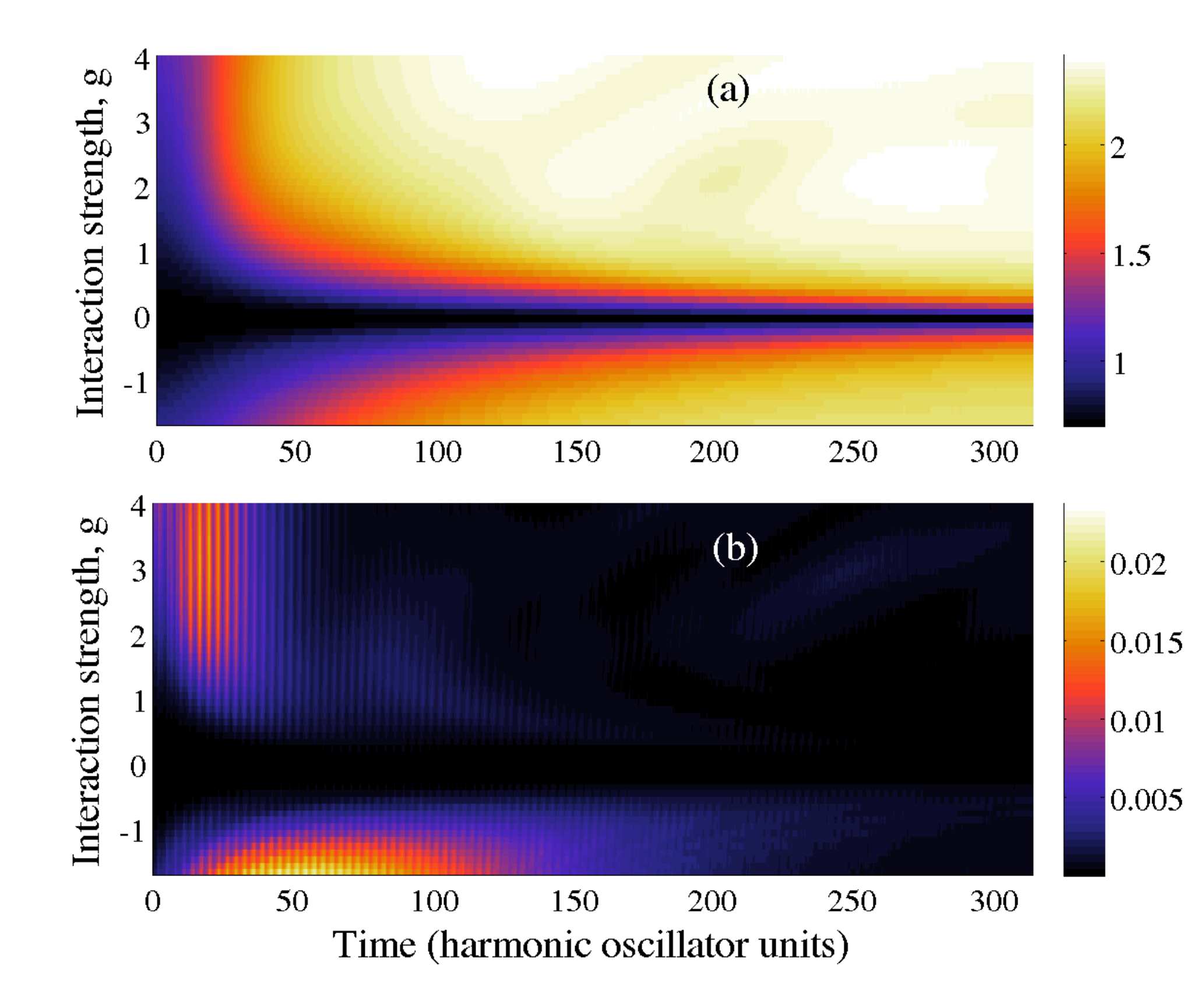}
\caption{(Color online): For $x_0 = 3$, (a) von Neumann entropy averaged over a time period of $2 \pi$, as defined by \refeq{eq:VNE} and \refeq{eq:VNEav} (b) the standard deviation of this quantity, given by the square-root of \refeq{eq:VNEvar}, for a range of interaction strengths both repulsive and attractive.  Entropy increases gradually at early times $t < 10 \pi$, then increases at a more rapid rate before leveling off to an almost constant value with small fluctuations.  This behavior is similar for both attractive and repulsive interactions.  The variance over the $2 \pi$ averaging range behaves very differently for strong attractive and repulsive interactions, with the short-timescale fluctuations persisting for much longer if $g<0$. This difference is explained by a change in the dominant processes, with the attractive system being unable to excite the relative degrees of freedom in a cluster and thus transfer of atoms between each cluster becoming more significant.  Fig. \ref{fig:expdisp} (b) shows atom transfer dynamics in the repulsive case have only small fluctuations at late times.
}

 \label{fig:VNE}
\end{figure}

One questions of interest is whether the system reaches an equilibrium at long times.  We attempt to quantify this by looking at the single body density matrix and its von Neumann entropy, given by \refeq{eq:VNE};  however, this quantity (like most in our system) has a time-dependence due to the repeated collisions that are a consequence of the system as a whole being held within a harmonic confining potential. 
In order to simplify our analysis we look at the time averaged value over a period of $T=2 \pi$ and quantify the degree of short-time change via the variance of this average.  These are plotted in \reffig{fig:VNE}; (a) shows that for both positive and negative $g$, $S_{\mathrm{VN}}$ increase towards a maximum value, with small amplitude oscillations in a similar way to $\Delta N$ but with much smaller variations. For fixed $|g|$, the $g>0$ entropy generally increases slightly faster and to higher values than the equivalent $g<0$ case, but is otherwise quite similar.  
Fig. b) shows the standard deviation over the $2 \pi$ averaging period, the rapidly changing (time scales of less than $2 \pi$) effects continue for much longer in the attractive case compared to the repulsive.  Transfer effects [discussed in Sec. \ref{subsubsec:transfer}] are likely the cause of this short time oscillation as they are predicted to be cyclic on the timescale of a few collisions when $g \approx 1$.  The variation dying down at long times can be explained for the $g>0$ case  by intra-cluster exited states breaking the cyclic effect, and for $g<0$, by the slower effect of the broadening of the inter-cluster wavefunction to the point where the collision time is not well defined.

\section{Conclusions \label{sec:conc}}

We have considered a system of $N=4$ atoms with contact interactions, confined within in a harmonic potential.  Our initial condition was a symmetric setup of two $N/2$ atom ground states, displaced from one another by a distance $x_0$ (taken to be 3 harmonic oscillator lengths for most of the numerics), which we then left to oscillate and undergo collisions.  Initially there is no entanglement between the atoms on the left and on the right, however interactions lead to the generation of entanglement. 

We investigated left/right number variation within the system, based on an operator which could in principle be measured directly in the experimental setup we suggest in this paper.  Initially both (left and right) states have a near definite number of two atoms and hence a number uncertainty $\Delta N_{\rm R}$, which is initially close to zero.  When the left and right states are well separated, $\Delta N_{\rm R}$ is a measure of entanglement between the left and right sides.  However when the two states are close, i.e., during collisions, $\Delta N_{\rm R}\sim N/4=1$; we therefore investigated the difference from minimum-to-minimum value taken over a time range of around $\pi$, 
i.e., the minimum value of $\Delta N_{\rm R}$ obtained after the $n$th collision.   
There is a marked difference in the evolution of $\Delta N_{\rm R}$ between the $g<0$ (attractive) and $g>0$ (repulsive) cases.  When $|g| \gtrapprox 0.5$, number uncertainty builds up much more slowly with attractive interactions than with repulsive, essentially resisting entanglement. This is despite a large increase to the change in number uncertainty that is generated by a single collision. This increases quadratically with $|g|$ when $g \lessapprox -1.3$, but in the repulsive case the increase reaches a maximum, and then drops as $g$ increases further.  Additionally for $g>0$ we observe long-timescale high-amplitude number fluctuations, which continue even at late times (over 100 collisions).  

This behavior is explained by our time dependent perturbation theory on the atom transfer process, and the energy difference between the intra-cluster excited states. We investigated the effect of $\Delta E_{\rm int}$, the energy difference in intra-cluster energies between the $\{ 2,2\}$ (two displaced $N=2$ ground states) and $\{3,1\}$ (one free atom and one $N=3$ atom ground state) configurations.  Assuming the average interaction energy between the clusters to be weak (i.e.\ $|Ag| \ll 1$), increases to $|\Delta E_{\rm int}|$ lead to a phase mismatch 
and thus to destructive interference so that the population transfer cycles periodically.  If intra-cluster excited states are present, this picture breaks down, since each of these excited states phase-evolves at a different rate; cancellation becomes more complicated and the states less localized, which occurs for large $g>0$ at long times.  
The energy gap between the ground and excited states of each of the $N/2$ atom clusters is increased (decreased)  when $g$ gets smaller (larger), 
which reduces the maximum population that can be transferred to excited states. The excited states become effectively inaccessible as $g \ll 0$, resulting in an effectively two-level system of the $\{2,2\}$ and $\{3,1\}$ configurations. 
Our perturbation theory indicates that for sufficiently strong attractive interactions, with very specific values,  phase matching would be possible, allowing for resonant transfer. However this is outside the regime our numerical method is capable of reliably portraying, and will remain an avenue for future research.  

By separating the system into components of the wavefunction with definite number (number states of the number-to-the-right operator) we have observed the evolution of the positions associated with one/two/three atom number states, and the right side position variance.  For $g=3$ the peaks in position variance increase to a maximum for all $N_R = n$ in around 100 harmonic time units ($100/2 \pi$ oscillator periods or around 30 collisions) and do not fluctuate greatly. Considering instead the case where $g = -1.7$, after $~60$ collisions, we find that for $N_R \ne 2$ position and position uncertainty are the same as they are for a state undergoing collision, whereas the $N_R = 2$ tends to a maximum. This indicates that the state is well described by two atomic dimers with a significant uncertainty in their relative displacement and almost no amplitude of a singlet-trimer like state is present in the wavefunction; this motion again undergoes partial revivals on very long timescales.    

In addition, we have investigated the von Neumann entropy of the single-body-density matrix $S_{\mathrm{VN}}(t)$, in order to investigate to what degree the system tends to an equilibrium.  We note $S_{\mathrm{VN}}(t)$ is zero for a product state (all atoms with the same wavefunction/occupying the same mode) and can be considered a measure of how mean-field-like the state is.  Additionally $S_{\mathrm{VN}}(t)$ is constant for our system if $g=0$, despite the wavefunction evolving periodically in time.  
At long times with repulsive interactions, $S_{\mathrm{VN}}$ (time averaged over a period of $2 \pi$) increases to a steady value with only small fluctuations over the averaging period. However, long-term fluctuations (over the order of twenty $\pi$ time units) are still present and appear to be due to atom transfer processes which do not appear to equilibrate on the timescales considered in this paper.  The time required to reach maximum entropy decreases with larger $g$ but this appears to saturate with little change for $g\gtrapprox 2$; for an initial separation of $x_0 =3$ this takes around 30 collisions. This short-term increase appears to be due to the inter-cluster degrees of freedom discussed in the previous paragraph; the associated probability density with the separation of the two clusters becomes less peaked.  
With very weak attractive interactions, the system's behavior is similar to the repulsive case, however for $|g|\gtrapprox 0.5$ higher intra-cluster excited states become less accessible, leading effectively to a reduction in the number of accessible degrees of freedom, such that the left/right states behave more like solitons.  In this case, the time average of $S_{\mathrm{VN}}(t)$ does not tend to a long-term mean value as 
compared with the case of repulsive interactions of similar magnitude; there is also a great deal more short-time variation, which persists for longer. The short-time variation  can be attributed to the strong atom transfer effects, which are predicted to cycle population continually due to an energy difference.  The effect eventually reduces as displacement uncertainty between the two bound states (which now behave like quantum solitons) increases, which is the mechanism behind the long term entropy increase.  

A pseudo-periodicity effect is also present. The non-interacting system is periodic with a period $\pi$, and thus the Fourier transform of any time dependent expectation values will have frequency peaks at $n/\pi$.  We have examined how these peaks shift for the left/right number uncertainty as interaction strength is varied and have found an approximately linear shift with $g$ over the range considered.  Changes to higher order components of the frequency spectrum depend deviate slightly from the linear dependence shown by the first order, with differences only clearly manifest for $|g|\gtrsim 1$).

\acknowledgments

We would like to thank the UK EPSRC for funding (Grant No. EP/G056781/1) and the Jack Dodd Centre (S.A.G.) for support, as well as Lincoln D. Carr for illuminating discussions.  

\begin{appendix}

\section{Identities involving Jacobi coordinates \label{app:Jacobi_proof}}
\subsection{First identity}
We wish to show that the Jacobi coordinates defined by \refeq{eq:xcdef} and \refeq{Eq:JacobiCoordinates} satisfy 
\begin{equation}
\sum_{k=1}^N x_k^2 = Nx_{\mathrm C(N)}^2 + \sum_{k=2}^N \frac{k-1}{k}\xi_{k}^2  \; .
\label{eq:sumxsqid2} 
\end{equation}
We prove this inductively. The $N=2$ case can readily be verified, after which we may consider the increase of number from $N-1$ to $N$.  In particular, 
\begin{equation}
\sum_{k=1}^N x_k^2  = x_{N}^{2} + (N-1)x_{\mathrm C(N-1)}^2 + \sum_{k=2}^{N-1} \frac{k-1}{k}\xi_{k}^2.
\end{equation}
Noting that $\xi_{N}=x_{N} - x_{\mathrm{C}(N-1)}$, we then deduce
\begin{equation}
\begin{split}
\sum_{k=1}^N x_k^2  =& x_{N}^{2} + (N-1)x_{\mathrm{C}(N-1)}^2 
\\&-
\frac{N-1}{N}\left[
x_{N} - x_{\mathrm{C}(N-1)}
\right]^{2}
+ \sum_{k=2}^{N} \frac{k-1}{k}\xi_{k}^2. 
\end{split}
\end{equation}
Collecting terms, this reduces to
\begin{equation}
\begin{split}
\sum_{k=1}^N x_k^2  = & 
\frac{1}{N}\left[
x_{N} + (N-1)x_{\mathrm{C}(N-1)} 
\right]^{2}
+ \sum_{k=2}^{N} \frac{k-1}{k}\xi_{k}^2 \\
=&
N x_{\mathrm{C}(N)}^{2} + \sum_{k=2}^{N} \frac{k-1}{k}\xi_{k}^2 \;, 
\end{split}
\end{equation}
which completes the proof.  An equivalent result also holds in 3D~\cite{Yamada2009}.

\subsection{Second identity}
We rephrase \refeq{Eq:JacobiCoordinates} as 
$
x_{k} = \xi_{k} + [1/(k-1)]\sum_{j=1}^{k-1} x_{j}
$.  Recursively substituting in equivalent expressions for $x_{k-1}, x_{k-2},\ldots, x_{N/2+1}$ yields (for $N/2+1<k\leq N$)
\begin{equation}
x_{k} = \xi_{k}+ \sum_{j=N/2+1}^{k-1}\frac{\xi_{j}}{j} + \frac{1}{N/2}\sum_{j=1}^{N/2}x_{j},
\end{equation}
and for $k=N/2+1$ we have
$x_{N/2+1} = \xi_{N/2+1} + (2/N)\sum_{j=1}^{N/2} x_{j}$. Hence, summing over all $k\in\{N/2+1,N/2+2,\ldots,N\}$,
\begin{equation}
\begin{split}
\sum_{k=N/2+1}^{N}x_{k}  = & 
\sum_{k=N/2+1}^{N}\xi_{k} +
\sum_{k=N/2+2}^{N}\sum_{j=N/2+1}^{k-1}\frac{\xi_{j}}{j} 
+  \sum_{k=1}^{N/2} x_{k}
\\
=& 
\sum_{k=N/2+1}^{N}\xi_{k} +
\sum_{k=N/2+1}^{N-1}\frac{N-k}{k} \xi_{k} 
+  \sum_{k=1}^{N/2} x_{k}
\\
=&
\sum_{k=N/2+1}^{N}\frac{N}{k} \xi_{k}
+  \sum_{k=1}^{N/2} x_{k} \;,
\end{split}
\end{equation} 
from which we deduce the desired identity:
\begin{equation}
\sum_{k=N/2+1}^{N}x_{k}  
- \sum_{k=1}^{N/2} x_{k}
=  \sum_{k=N/2+1}^{N}\frac{N}{k} \xi_{k}.
\label{eq:secondJidentity}
\end{equation}

\section{Calculations for the number-to-the-right operator\label{app:num_op}}

\subsection{Analytically determined properties of $\hat{N}_R^2$ \label{app:AnalyticLRvar}} 
From the definition of \refeq{eq:numlr}, it follows that
\begin{align}
 \hat{N}_R^2 &=  \int_{0}^{\infty} \mathop{dx dx'}  \hat{\Psi}^{\dagger} (x)  \hat{\Psi}^{\dagger} (x')  \hat{\Psi}(x) \hat{\Psi}(x') + \hat{N}_R \; ,
\end{align}
and, given a general (symmetrized) many-body wavefunction $\psi(\vec{x})$, one may deduce the expectation values
\begin{align}
 \ev{\hat{N}_R} = & N \int_{0}^{\infty} \mathop{dx_{1}} \int_{-\infty}^{\infty} \mathop{dx_{2}\ldots dx_N} \vert \psi(\vec{x})\vert^2 \;, \\
\ev{\hat{N}_R^2 }  =&  N(N-1)\int_{0}^{\infty} \mathop{dx_{1} dx_{2}}  \int_{-\infty}^{\infty} \mathop{dx_3
\ldots
dx_N} 
\vert \psi(\vec{x})\vert^2 
+\ev{\hat{N}_R} \;.
\end{align}
For a product-state wavefunction $\psi(\vec{x}) = \prod_{k=1}^{N} \phi(x_k)$, expectation values are simple to calculate, as all integrals are separable and most evaluate to unity.  In this case\begin{align}
 \ev{\hat{N}_R} &= N \int_0^{\infty} \mathop{dx} \vert \phi(x) \vert^2 \;, \\
\begin{split}
   \ev{\hat{N}_R^2} &= N(N-1)  \left[ \int_0^{\infty} \mathop{dx} \vert \phi(x) \vert^2  \right]^2 +  \ev{\hat{N}_R}  \\
		  &= [(N-1)/N]  \ev{\hat{N}_R}^2  +  \ev{\hat{N}_R}  \;,
\end{split}
\end{align}
and so the variance of $\hat{N}_{R}$ for a product state simplifies to
\begin{equation}
  \Delta_{P} N_{R}
   = \ev{\hat{N}_R} (1-\ev{\hat{N}_R}/N) \;.
\label{eq:DeltaNR}
\end{equation}

We may determine analytic expressions when $g=0$, which, for the purpose of this paper, we limit to the $N=4$ case.  Without interactions, our many body wavefunction is given by \refeq{eq:initconNI}, and
\begin{gather}
\int_0^{\infty} dx |\phi(x,\pm x_0,t)|^2 = \frac{1}{2}\left[1\pm \mbox{erf}(x_0 \cos(t)) \right] \;,  
\\
\int_{-\infty}^{\infty} dx \phi^{*}(x,\pm x_0,t) \phi(x,\mp x_0,t)= \mathrm{e}^{-x_0^2 \pm i x_0 \sin(2t)/2} \;,  \\
\begin{split}
\int_0^{\infty} dx \phi^{*}(x,\pm x_0,t) \phi(x,\mp x_0,t) = 
&\frac{1}{2} 
\left[1 \pm \mbox{erf}(x_0 \sin(t)) \right]
 \\ &\times
\mathrm{e}^{-x_0^2 \pm i x_0 \sin(2t)/2}
 \;,
\end{split}
\end{gather}
with erf denoting the error function.  Calculating $\ev{\hat{N}_R^2}$ in principle requires accounting for 36 different terms, however, assuming we can neglect terms proportional to $\exp(-2x_0^2)$, only 6 are important, 
and we have
\begin{equation}
\begin{split}
  \ev{\hat{N}_R^2} 
  \approx & \frac{N(N-1)}{24} \left\{ [1-\mbox{erf}(x_0 \cos(t))]^2 \right.   \\
 & +  4[1-\mbox{erf}^2(x_0 \cos(t))] 
 \\ 
  & + \left. [1+\mbox{erf}(x_0 \cos(t))]^2\right\} 
  +\ev{\hat{N}_R} \\
  = &5 -\mbox{erf}^2[x_0 \cos(t)]\;.
\end{split}
\end{equation}
Subtracting 4 then yields the variance as given by \refeq{eq:DeltaNRnistate}.

\subsection{Numerical calculation of number variance}
In order to calculate the number variance we decompose the field operator into our basis set, $\hat{\Psi}(x) = \sum_k \hat{a}_k \phi_k(x)$.  In this form we can express $\hat{N}_R^2$ as
\begin{align}
\hat{N}_R^2  = \sum_{i,j,k,\ell} y_{ik} y_{j\ell} \hat{a}^{\dagger}_i \hat{a}^{\dagger}_j  \hat{a}_k \hat{a}_{\ell} + \hat{N}_R   \; ,
\end{align}
where $y_{j \ell} = \int_0^{\infty} \mathop{dx} \varphi_j(x) \varphi_{\ell}(x) $ is the positive space overlap between two Hermite functions, given by $\delta_{j \ell}/2 $ if $j+\ell$ is even, and otherwise given by
\begin{multline}
y_{j \ell} =
(-1)^{(j+\ell-1)/2}  \phantom{F}_2 F_1( -j,1-[j-\ell]/2;1-[j+\ell]/2,-1 )  \\
\times \frac{2^{-j} (j+\ell+2)!!}{\sqrt{2 \pi j! \ell!}} \;, 
\end{multline}
where $\phantom{F}_2F_1$ denotes a standard hypergeometric function. Likewise the integral from minus infinity to zero is $(-1)^{j+\ell} y_{j \ell}$. This formula is useful for small numbers and testing, but for practical purposes we calculate the integral via Gauss Laguerre quadrature, which is numerically exact for odd $j+\ell$ (all other cases are trivially zero or one half) given a rule of order $(j+\ell+1)/2$ or higher.  Given our truncated basis and symmetry about $x=0$, this can be expressed as a finite size matrix of only even-parity functions with $\ev{\hat{N}_R}=N/2$ just a numerical constant for our initial condition. 

\subsection{Numerical calculation of restricted region expectation values}
In addition to this we wish to calculate expectation values in restricted regions via \refeq{eq:restricted_ev}, corresponding to sections of the wavefunction with exactly $n$ particles to the left or right, along with the associated normalization factors when the wavefunction is divided into these regions.  If our many body wavefunction is $\psi(\vec{x})$ then the normalization factors are given by
\begin{align}
{\cal N}_n = \frac{N!}{(N-n)!n!} \int_0^{\infty} \mathop{dx_1\ldots dx_n} \int_{-\infty}^0 \mathop{dx_{n+1}\ldots dx_{N}} \vert \psi(\vec{x}) \vert^2 \; , 
\end{align}
and the expectation value of the distance to the right operator is equal to 
\begin{equation}
\begin{split}
\ev{\hat{x}^{(n)}_{\rm R}} = & {\cal N}_n^{-1} \int_{-\infty}^{\infty} \mathop{dx_1\ldots dx_{N}} 
\sum_{k=0}^N x_k \theta(x_k)  \\
 & \times \sum_{\cal P}  \prod_{k=1}^n \Theta(x_k) \prod_{j=n+1}^N \Theta(-x_j)  \mathop{\vert \psi(\vec{x}) \vert^2}    \\
= &{\cal N}_n^{-1}\frac{N!}{(N-n)!n!} 
\int_0^{\infty} \mathop{dx_1\ldots dx_{n}} 
\\&\times 
\int_{-\infty}^0 \mathop{dx_{n+1}\ldots dx_{N}} \sum_{k=n+1}^N x_k \mathop{\vert \psi(\vec{x}) \vert^2} \; . 
\end{split}
\end{equation}
For computation, these operators are converted into matrix form by taking the matrix elements between different elements of the basis set, and then projected to our reduced (center-of-mass ground state) basis. 

\section{Two cluster wavefunction evolution \label{app:2_cluster_coms}}
Here we derive the time dependent wavefunction describing the center of masses of our two cluster system, i.e. the part acted on by $\hat{H}_{L/R}^{(\rm C)}$, the center-of-mass components from \refeq{eq:split_ham}; with $\hat{H}_{I}$ ignored. Denoting $y_1,y_2$ as the coordinates of the center-of-masses of each cluster, up to a normalization factor our initial two-cluster wavefunction is given by  
\begin{align}
\bket{y_1,y_2}{\varphi_{n,N-n}(0)} \propto &\exp\left(-\frac{N-n}{2}\left[y_2 + \frac{n X_n}{N- n}\right]^2\right) \nonumber \\
&\times \exp\left(-\frac{n}{2}[y_1 - X_n]^2\right)+ {\rm perm} \; ,
\label{eq:two_cluster_init}
\end{align}
with ``perm'' indicating the term obtained by permuting $y_1$ and $y_2$, as required by symmetry. This gives rise to a time-dependent normalization constant which we do not discuss here.  If we instead express this in terms of $y_{\rm C} = [n y_1 + (N-n) y_2]/N$ and $y_{\rm R} = y_1-y_2$ we have
\begin{align}
\bket{y_{\rm C},y_{\rm R}}{\varphi_{n,N-n}(0)}\propto &\exp\left(-\frac{n [(N-n) y_{\rm R} -N X_{n}]^2}{2N[N-n]}\right) \nonumber \\
&\times \exp\left(-\frac{N y_{\rm C}^2}{2} \right) + {\rm perm} \; ,
\label{eq:two_cluster_init_2}
\end{align}
where in this case ``perm'' is simply flipping the sign of $y_{\rm R}$, and we can factor out the $y_{\rm C}$ dependence.  If we temporarily ignore interactions between the two clusters, it is straightforward to generalize this to the time dependent case via \refeq{eq:osccoh}:
\begin{multline}
\bket{y_{\rm C},y_{\rm R}}{\varphi_{n,N-n}(t)}\propto \\
\exp\left(-\frac{n [(N-n) y_{\rm R} -N X_{n} \cos(t)]^2}{2N[N-n]}\right) 
\exp\left(-\frac{Ny_{\rm C}^2}{2} \right)
\\
\times
 \exp\left(i\left[t- n y_{\rm r} X_n \sin(t)  +
  \frac{X_n}{4} \left(n-\frac{n}{N-n}\right)\sin(2t)\right]\right)
  \\ + {\rm perm} \; .
\label{eq:two_cluster_osc}
\end{multline}
Interactions between clusters can modify only the $y_{\rm R}$ dependent part of this wavefunction. %

\section{Energy bound for Hamiltonian variance \label{app:energy_bound} }

As the Hamiltonian is time independent, the time evolution operator commutes with all powers of the Hamiltonian. Denoting our state as $| \psi(t)\rangle$ we have for any time $t$
\begin{equation}
\langle \psi(t)|\hat{H}^n| \psi(t)\rangle = \langle \psi(0)|\hat{H}^n| \psi(0)\rangle\;,\quad n=1,2,\ldots \; .
\end{equation}
As absolute values of energy are not physically important, we consider a re-zeroed Hamiltonian
\begin{equation}
\hat{\cal H}=  \hat{H} - \langle \psi(0)|\hat{H}| \psi(0)\rangle \;,
\end{equation}
as it will make the mathematics more convenient.  Introducing the notation for the variance of the re-zeroed Hamiltonian 
\begin{align}
\Delta E^2 =  \langle \hat{\cal H}^2 \rangle\; ,    
\end{align}
we note that this quantity must be positive and real as $\hat{H}$ is a Hermitian operator. 

Let us define two wave functions $|\psi_1(t)\rangle$ and $|\psi_2(t)\rangle$ as being negligibly mixed at a certain point in time if
\begin{equation}
\label{eq:M}
\langle\psi_1(t)|\hat{\cal H}^n |\psi_2(t)\rangle \le \eta \;,\quad n=1,2
\end{equation}
with $\eta$ a small parameter.  Note that in lattice models $\eta$ could be exactly zero up to some finite power $n$. 
If both the initial wave function and $|\psi_{1,2}(t)\rangle$ are normalized to one and the latter are negligibly mixed, the wave function at time $t$ can be written (up to a global phase factor) as
\begin{equation}
|\psi(t)\rangle = \sqrt{p}|\psi_1(t)\rangle + \sqrt{1-p}e^{i\alpha} |\psi_2(t)\rangle
\end{equation}
with real $\alpha$ and $0\le p\le 1$.  
Introducing the notation 
\begin{align}
\langle \hat{\cal H}^n \rangle_{j}&\equiv\langle \psi_{j}(t)|\hat{\cal H}^n| \psi_{j}(t)\rangle\;,
\end{align} 
we can see from \refeq{eq:M} and the fact that the expectation value of total Hamiltonian is zero, that these two quantities are related via
\begin{align}
\langle \hat{\cal H} \rangle_{1} =  \frac{p-1}{p}\langle \hat{\cal H} \rangle_{2}  +O(\eta) \;.
\label{eq:1&2}
\end{align} 
Setting $\eta=0$ in \refeq{eq:M} we have for $n=2$
\begin{align}
\Delta E^2 &= p\langle \hat{\cal H}^2 \rangle_{1} + (1-p)\langle \hat{\cal H}^2 \rangle_{2} \nonumber \\
	   &\ge  p\langle \hat{\cal H} \rangle_{1}^2 + (1-p)\langle \hat{\cal H} \rangle_{2}^2 \;,
\end{align} 
with the second step true again by the fact that $\hat{H}$ is Hermitian.  Finally, substituting in for $\langle  \hat{\cal H}  \rangle_{1}$ via \refeq{eq:1&2} we obtain
\begin{align}
\Delta E^2 &\ge \frac{1-p}{p} \langle \hat{\cal H} \rangle_{2}^2 \;, \\
\Delta E^2 &\ge \frac{p}{1-p} \langle \hat{\cal H} \rangle_{1}^2 \;, \\
\Delta E^2 &\ge  \frac{\left( \langle \hat{\cal H} \rangle_{1} -\langle \hat{\cal H} \rangle_{2} \right)^2}{p(1-p)}   \;, 
\end{align}
which leads to \refeq{eq:indv_state_bounds} in the main text.

\subsection{Analytic calculations of $\Delta E$ \label{app:deltaE}}

For our two particle initial condition, if $x_0 \gg 1$, i.e., well-separated initial clusters, we can analytically determine $E$ and $\Delta E$. Within this well-separated approximation we only need to consider one cluster, displaced a distance $x_0$ from the center, and multiply by 2 to get the values for the whole wavefunction.   For dimers, our wavefunction is $f(x_1-x_0,x_2-x_0)^{(2)}$ as defined in \refeq{eq:twobodygs}, otherwise it is not analytic.   This wavefunction is still an eigenstate of the relative Hamiltonian (for $n$ particles), with some eigenvalue $E_{\rm rel}^{(n)}$, but not of the center-of-mass part. Therefore we need only consider the center-of-mass Hamiltonian
\begin{equation}
 H_{\rm C}(x_{\mathrm{C}}) = -\frac{1}{2n} \pdsq{x_{\rm C}} + \frac{n x_{\rm C}^2}{2} \;,
\end{equation}
acting on the displaced ground state
\begin{equation}
 \psi_{\rm C} (x_{\rm C}) = \left(\frac{n}{\pi}\right)^{1/4}\exp( -n[x_{\rm C} -x_0 ]/2) \;,
\end{equation}
to get all contributions to the variance.  Acting the Hamiltonian on this wavefunction we obtain
\begin{align}
 H_{\rm C} \psi_{\rm C} (x_{\rm C}) &= \left(\frac{1}{2} + n x_0 x + \frac{n x_0^2}{2} \right)\psi_{\rm C} (x_{\rm C}) \;, \\ 
H_{\rm C}^2 \psi_{\rm C} (x_{\rm C})&= \left[\frac{1}{4} + \frac{n x_0}{2}\left(4x - 3x_0 \right) + n^2 x_0^2(x_0-2x)^2 \right]\psi_{\rm C}  (x_{\rm C}) \nonumber  \;,
\end{align}
which can then be used to determine the expectation values
\begin{equation}
\begin{split}
 \ev{ \hat{H}_{\rm C} }&= \frac{1}{2} + \frac{nx_0^2}{2} \;, \\ 
\ev{ \hat{H}_{\rm C}^2 } &= \frac{1}{4} + nx_0^2 +\frac{n^2 x_0^4}{4}  \;.
\end{split}
\end{equation}
$\Delta E$ can then be calculated as the standard deviation of two times $\hat{H}_{\rm C} $
\begin{align}
\Delta E = 2\sqrt{\ev{\hat{H}_{\rm C} ^2}-\ev{\hat{H}_{\rm C} }^2} =  \sqrt{2 n} x_0 \;,
\label{eq:DeltEanal}
\end{align}
which is twice the square root of the difference between the initial (dimensionless) potential energy and the ground state energy. The reasons for this are similar to why a classical coherent state with an average value of $N$ photons has a shot noise proportional to $N^{1/2}$.  
Note that this result relies on $\exp(-n x_0^2) \ll 1$ and so can only be considered valid to this order.

\end{appendix}

\end{document}